\documentclass{aa}
\usepackage{array}
\usepackage{graphicx}
\usepackage{multirow}
\usepackage{color}
\usepackage{xcolor}
\usepackage{hyperref}
\usepackage{capt-of}
\usepackage{comment}
\usepackage[capitalize]{cleveref}
\usepackage{placeins}
\usepackage{txfonts}
\usepackage{makecell}
\renewcommand{\arraystretch}{1.2}

\def\N1316{NGC\,1316}
\def\N5204{NGC\,5204}

\def\xmm{XMM-{\textit{Newton}} }

\def\aap{{A\&A}}

\def\apjl{{ApJ}}

\def\apjs{{ApJS}}

\def\arcsec{\ifmmode '' \else $''$\fi}
\def\arcmin{$'$}
\def\arcsecpoint{\ifmmode ''\!. \else $''\!.$\fi}

\def\kms{\ifmmode {\rm km\ s}^{-1} \else km s$^{-1}$\fi}
\def\Msun{\ifmmode {\rm M}_{\odot} \else M$_{\odot}$\fi}
\def\Lsun{\ifmmode {\rm L}_{\odot} \else L$_{\odot}$\fi}
\def\Zsun{\ifmmode {\rm Z}_{\odot} \else Z$_{\odot}$\fi}

\def\ergscm2{ergs\,s$^{-1}$\,cm$^{-2}$}
\def\icm3{{\rm cm}^{-3}}
\def\icm2{{\rm cm}^{-2}}
\def\qo{\ifmmode q_{\rm o} \else $q_{\rm o}$\fi}
\def\Ho{\ifmmode H_{\rm o} \else $H_{\rm o}$\fi}
\def\ho{\ifmmode h_{\rm o} \else $h_{\rm o}$\fi}

\def\vFWHM{\ifmmode v_{\mbox{\tiny FWHM}} \else
            $v_{\mbox{\tiny FWHM}}$\fi}
\def\CCF{\ifmmode F_{\it CCF} \else $F_{\it CCF}$\fi}
\def\ACF{\ifmmode F_{\it ACF} \else $F_{\it ACF}$\fi}
\def\Halpha{\ifmmode {\rm H}\alpha \else H$\alpha$\fi}
\def\Hbeta{\ifmmode {\rm H}\beta \else H$\beta$\fi}
\def\Hgamma{\ifmmode {\rm H}\gamma \else H$\gamma$\fi}
\def\Hdelta{\ifmmode {\rm H}\delta \else H$\delta$\fi}
\def\Lya{\ifmmode {\rm Ly}\alpha \else Ly$\alpha$\fi}
\def\Lyb{\ifmmode {\rm Ly}\beta \else Ly$\beta$\fi}
\def\Lyg{\ifmmode {\rm Ly}\beta \else Ly$\gamma$\fi}

\def\ciii{\ifmmode {\rm C}\,{\sc iii} \else C\,{\sc iii}\fi}
\def\civ{\ifmmode {\rm C}\,{\sc iv} \else C\,{\sc iv}\fi}
\def\cv{\ifmmode {\rm C}\,{\sc v} \else C\,{\sc v}\fi}
\def\cvi{\ifmmode {\rm C}\,{\sc vi} \else C\,{\sc vi}\fi}

\def\o5007{[O\,{\sc iii}]\,$\lambda5007$}

\def\fexxii-iii{Fe\,{\sc xxii-xxiii}}

\usepackage{hyperref}
\usepackage{natbib}
\usepackage{stfloats}

\begin{document}

\title{Unveiling the biconical geometry of the outflow in the ultraluminous X-ray source NGC 5204 X-1}

%\subtitle{-}

\titlerunning{Unveiling the biconical geometry of the outflow in the ultraluminous X-ray source NGC 5204 X-1}
%%Authors
\author{S. Caserta\inst{1,2}\fnmsep\thanks{simona.caserta@unipa.it}
        \and
        C. Pinto\inst{2}
        \and
        T. Di Salvo\inst{1}
        \and
        F. Pintore\inst{2}
        \and
        P. Kosec\inst{3}
        \and F. Barra\inst{1,2}
        \and
        \\
          D. J. Walton\inst{4}
          \and
          A. D'Aì\inst{2}
          \and
          M. Del Santo\inst{2}
          \and
          A. Gúrpide\inst{5}
          \and
          A. Fabian\inst{6}
          \and
          A. Wolter\inst{7}
        }

\institute{Universit\`a degli Studi di Palermo, Dipartimento di Fisica e Chimica, via Archirafi 36, I-90123 Palermo, Italy
        \and
        INAF -- IASF Palermo, Via U. La Malfa 153, I-90146 Palermo, Italy
        \and
Center for Astrophysics | Harvard  \& Smithsonian, 60 Garden Street, Cambridge, MA 02138, USA
\and
            Centre for Astrophysics Research, University of Hertfordshire, College Lane, Hatfield AL10 9AB, UK
\and 
School of Physics \& Astronomy, University of Southampton, Southampton, Southampton SO17 1BJ, UK
\and
Institute of Astronomy, University of Cambridge, Madingley Road, Cambridge CB3 0HA, UK
\and
 INAF -- Osservatorio Astronomico di Brera, via Brera 28, I-20121 Milano, Italy
}

\date{XXXX; accepted YYYY}

\abstract
    {Ultraluminous X-ray sources (ULXs) are non-nuclear X-ray binary systems that exceed the Eddington luminosity for a $10$ M$_{\odot}$ black hole. The majority of these sources are thought to be stellar-mass compact objects accreting at super-Eddington rates, exhibiting powerful relativistic winds. These winds have been identified through the detection of absorption lines with a blueshift as high as $0.3c$ and emission lines typically found at their laboratory wavelengths.}
  % context heading (optional)
   {In this work, we analysed the XMM-\textit{Newton} data of the ULX NGC 5204 X-1, which has been observed to exhibit emission lines with a blueshift of about $0.3c$. The aim of this study is to examine the geometry and physical properties of the accretion disc and the relativistic outflows. In addition, we aim to explore the factors that influence the ULX spectral transitions.}
  % aims heading (mandatory)
   {We undertook an observing campaign with XMM-\textit{Newton} to explore the source behaviour at different luminosities. In this first paper of the series, we performed high-resolution X-ray spectroscopy, including archival data, with the RGS instrument which allowed us to resolve both emission and absorption lines. The outflows features were characterised using physical models of plasma in collisional-ionisation and photoionisation equilibrium.}
  % methods heading (mandatory)
   {We identify collisionally-ionised blueshifted and redshifted components at about $0.3c$. These findings have high statistical significance and suggest a biconical structure for the outflow. Additionally, the analysis of the O VII line triplet observed in the spectrum enables us to infer physical properties of the low-velocity line-emitting plasma, e.g. electron density ($n_\mathrm{e}\sim10^{10} \mathrm{cm}^{-3}$) and temperature ($T_e\geq1.5\times10^5$ K). A hybrid plasma whose ionisation balance is affected by both collisions and radiation is favoured.}
  % results heading (mandatory)
   {}
   {}

\maketitle

\section{Introduction} \label{sec:intro}

Ultraluminous X-ray sources (ULXs) are bright, mainly extragalactic, non-nuclear, point-like sources, emitting in the X-ray band with luminosities above $10^{39}$ erg/s, which corresponds to the Eddington limit for a 10 $\Msun$ black hole, assuming a steady and spherical emission (for recent reviews, see \citealt{King2023,Pinto2023a}).
The X-ray emission stems from the accretion of matter onto a compact object; however, their nature has been highly debated with two main possible scenarios.
They can be powered by intermediate-mass black holes (IMBHs, $10^{2-4}$ M$_{\odot}$; \citealt{imbhs}), accreting at sub-Eddington rates, or by compact objects with smaller masses, such as neutron stars (NSs) and stellar-mass black holes (BHs), accreting at super-Eddington rates. 
This latter scenario is supported by the discovery of coherent pulsations in numerous sources (the first being M82 X-2, \citealt{m82x2}), unambiguously confirming the accreting compact object to be a NS, and in turn revealing highly super-Eddington luminosities.
Additionally, ULX spectra differ from those of the sub-Eddington Galactic black hole X-ray binaries (XRBs): the former are characterised by a soft excess, usually modelled by a low-temperature blackbody component, plus a second thermal component with a curvature above $2\,\mathrm{keV}$, which might originate from a supercritical accretion disc \citep{ulxstate, Bachetti2013}. Additionally, evidence has emerged for a third, harder continuum component, revealed by high-statistics \textit{NuSTAR} data \citep{Walton2018}. Its origin is thought to be related to photon scattering either in an optically thin corona or within the accretion column onto a magnetised neutron star \citep{brightman2016,Walton2018}.

The theory of super-Eddington accretion predicts an accretion disc around the compact object which is geometrically and optically thick inside the spherisation radius, and within this radius an optically thick outflow, with a clumpy nature, is launched from the top of the disc by the radiation pressure (e.g. \citealt{supercritical_accr, Poutanen2007,takeuchi2013}). Due to the resulting funnel shape of the accretion disc/wind combination, the spectral hardness of the source below 10 keV -- typically defined for XMM-\textit{Newton} data as the ratio between the flux in the $2-10$ keV and in the $0.3-2$ keV energy bands -- depends on the viewing angle (e.g. \citealt{Middleton2015a}): at low inclinations from the rotational axis, the spectrum will appear hard (hard-ultraluminous regime or HUL), as we are able to see the innermost and hotter regions of the accretion disc; at higher angles, the hard X-ray emission is obscured by the disc and outflow, resulting in a softer spectrum (soft-ultraluminous regime or SUL, e.g. \citealt{uls}).

Several spectral features are observed in the soft energy band, below 2 keV, which have been identified thanks to the high-resolution grating spectrometers onboard XMM-\textit{Newton} and \textit{Chandra} (e.g. \citealt{spectral_lines1,spectral_lines2,pulx_swift}). Evidence for corresponding features has also been found in CCD data in most cases with high-counts spectra (see e.g. \citealt{Stobbart2006,1313x1_mid}) and in a few cases even in the Fe\,K band ($6-9$\,keV, e.g. \citealt{Walton2016,Brightman2022}). Their detection is important, as they reveal information on the geometry of the accretion disc and on the presence of outflows driven by the radiation pressure, which supports the scenario of super-Eddington accretion.
Mildly-relativistic outflows have been detected in many sources (see the first catalogue of spectral lines in a sample of ULXs by \citealt{Kosec2021}).

\section{NGC 5204 X-1} \label{sec:intro_target}
NGC 5204 X-1 is a bright and persistent ULX located at a distance of 4.5 Mpc \citep{distance}, with an observed (average) X-ray luminosity $L_\mathrm{X}\sim 4.5 \times 10^{39}$ erg/s \citep{luminosity_5204} in the $0.3-10$ keV energy range. The source mostly switches between hard-intermediate (HUL), soft-bright (SUL), and supersoft-faint (SSUL) X-ray regimes.
Furthermore, it shows a putative quasi-periodic flux modulation over a timescale of about 200 days \citep{gurpide}, likely due to a variation in the mass accretion rate or the system geometry.
NGC 5204 X-1 shows evidence of an ultrafast outflow (UFO) with a velocity of about $-0.3c$, revealed by the detection of blueshifted emission lines, i.e. the fastest outflow ever recorded in a ULX.
Specifically, it is the only ULX in which a UFO has been detected in emission; in other ULXs, such as NGC 1313 X-1 \citep{1313x1_pi} and the pulsating NGC 300 ULX-1 \citep{Kosec2018b}, UFOs have instead been detected in absorption.
This highly blueshifted emitter was modelled assuming the plasma in collisional-ionisation equilibrium by \citet{Kosec2018a} with a cumulative significance of $3\sigma$. They also find similarities with the source SS433, the only known persistent super-Eddington accretor in our Galaxy, in which relativistic blue- and redshifted emission lines from collisionally-ionised jets have been detected \citep{ss433_lines}.

The primary goals of this work are (i) to investigate the spectral transitions observed in ULXs, which might be driven by the wind, the accretion rate or the precession of the inner accretion flow and (ii) to constrain the structure of the outflow in analogy with the Galactic source SS433.
To this aim, we used XMM-\textit{Newton} data, combining archival data with two newly acquired long-exposure observations.

This paper is organised as follows: in Section \ref{sec:data_reduction}, we describe the available observations and the data reduction procedure; Section \ref{sec:analysis} presents the spectral continuum modelling and the different methods used to investigate the outflows; in Sections \ref{sec:discussion} and \ref{sec:conclusions}, we discuss and summarise our results.

\section{Observations and data reduction} \label{sec:data_reduction}
We used data from all available XMM-\textit{Newton} observations of \N5204 X-1 between 2003 and 2023 (see Table \ref{table:observations}), including the last 240 ks campaign (ObsIDs 0921360101 and 0921360201; PI: Pinto).
The latter was split into two epochs separated by 6 months during 2023, in order to investigate the source in two different flux regimes. Through a \textit{Swift} X-ray telescope (XRT) \citep{swift} monitoring, we triggered two observations. The first was obtained when the source was in the unobscured soft-bright regime (ObsID 0921360201), specifically when the source XRT count rate was above $0.07$ counts/s and the hardness (calculated as the ratio between the counts in $1.5-10\,\mathrm{keV}$ and $0.3-10\,\mathrm{keV}$ bands) was below 0.4. The second observation was triggered when the source was in the hard-intermediate regime (ObsID 0921360101), when the count rate was between $0.05 -0.07$ counts/s and the hardness above 0.4. These flux regimes are representative of previously observed states of the source, as XMM-\textit{Newton} has observed NGC 5204 X-1 in both the soft-bright regime (EPIC-pn count rates $>0.7$) and the hard-intermediate regime (EPIC-pn count rates $\sim0.5$; see Table \ref{table:observations}).

\begin{table}[ht]
\centering
\caption{Table of the \xmm observations of \N5204 X-1.}
\label{table:observations}  
\renewcommand{\arraystretch}{1.3}
 \small\addtolength{\tabcolsep}{0pt}
 \vspace{-0.1cm}
\scalebox{0.95}{%
\hskip-0.0cm\begin{tabular}{@{}ccccccc}  
\hline
ObsID  &  Date & t$_{\text{RGS}1,2}$ & t$_{\text{pn}}$ & t$_{\text{MOS1}}$ & t$_{\text{MOS2}}$ & CR$_{\mathrm{pn}}$ \\
 &        &  (ks)  & (ks) & (ks) & (ks) & (counts/s)\\
\hline
\multicolumn{7}{c}{\textbf{Archival observations}} \\
\hline
 0142770101 & 2003-01-06 & 10  & 15  & 18 & 18 & $0.515 \pm 0.006$\\
 0142770301 & 2003-04-25 & 18  & 3.5 & 6.4 & 6.4 & $0.720 \pm 0.014$\\
 0150650301 & 2003-05-01 & 11  & 4.8 & 7.7 & 7.9 & $0.846\pm0.013$\\
 0405690101 & 2006-11-15 & 28  & 7.7  & 13 & 13 & $1.069\pm0.012$\\
 0405690201 & 2006-11-19 & 45  & 28  & 42 & 44 & $0.874\pm0.006$\\
 0405690501 & 2006-11-25 & 42  & 20  & 31 & 31 & $0.646\pm0.006$\\
 0693851401 & 2013-04-21 & 17  & 13  & 16 & 16 & $0.507\pm0.006$\\
 0693850701 & 2013-04-29 & 17  & 10  & 16 & 16 & $0.530\pm0.007$ \\
 0741960101 & 2014-06-27 & 23  & 19  & 23 & 23 & $0.489\pm0.005$\\
\hline
\multicolumn{7}{c}{\textbf{New observations}} \\
\hline
0921360101 & 2023-05-18 & 114  & 98  & 113 & 113 & $0.589\pm0.002$\\
 0921360201 & 2023-11-10 & 95 & 36  & 68 & 68 & $0.779\pm0.005$\\
\hline
\end{tabular}}
\vspace{0.2cm}

Notes: Exposure times (t$_{\text{RGS}1,2}$, t$_{\text{pn}}$, t$_{\text{MOS1}}$, t$_{\text{MOS2}}$) account for the removal of periods of high background rate. $\mathrm{CR}_{\mathrm{pn}}$ are the net source count rates in the $0.3 - 10$ keV energy band.
\vspace{-0.2cm}
\end{table}

\begin{figure*}
   \centering
   \includegraphics[scale=0.7]{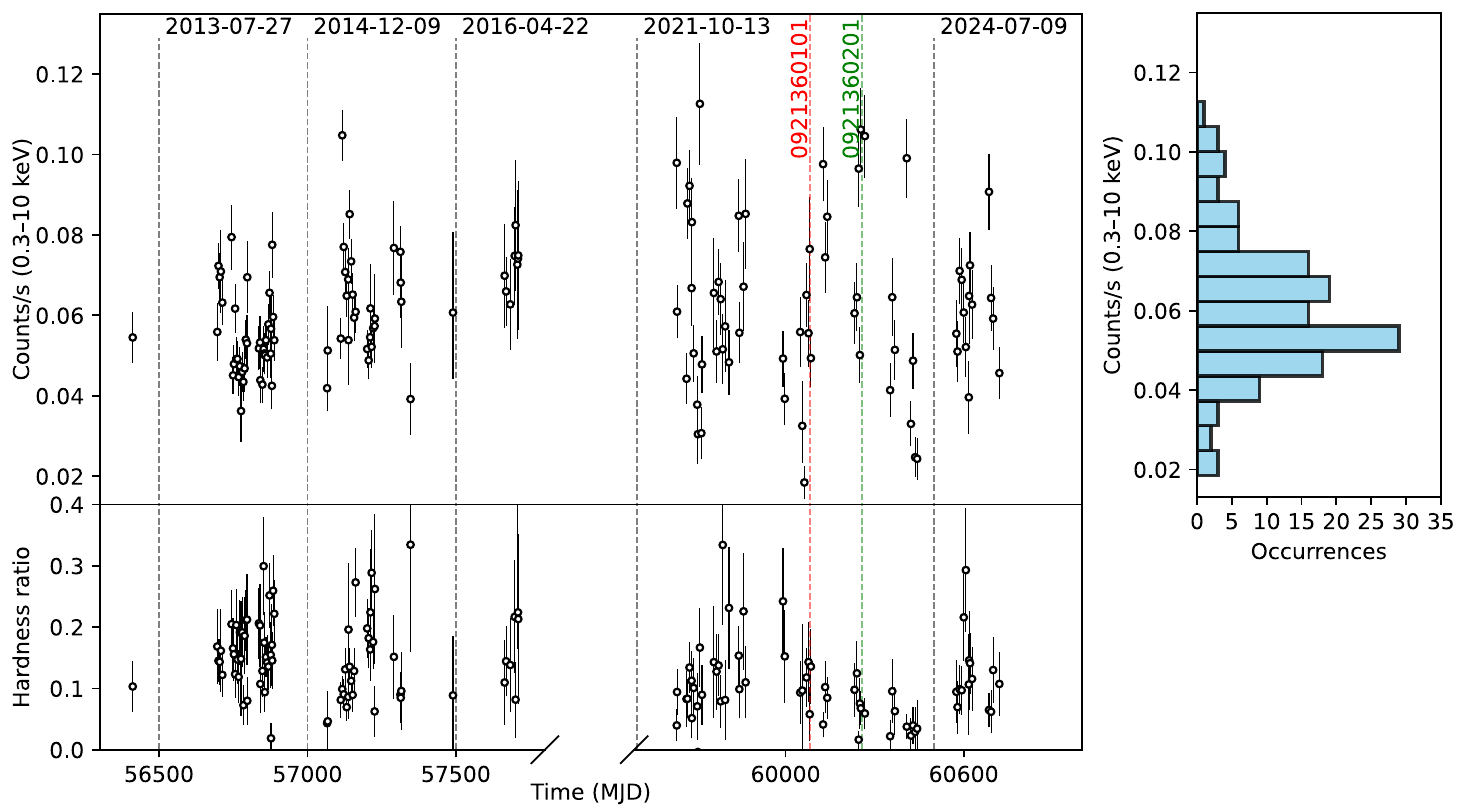}
   \vspace{-0.2cm}
   \caption{Left: \textit{Swift}/XRT light curve (top) and hardness ratio curve (bottom) of NGC 5204 X-1 from April 2013 to February 2025. The hardness ratio is defined as the ratio between the XRT counts in $3-10$ keV range and those in $0.3-3$ keV. The red and green dashed lines indicate the start times of the two latest XMM-\textit{Newton} observations. Right: Corresponding count rate histogram.}
    \label{fig:xrt_lightcurve}
   \vspace{-0.1cm} 
\end{figure*}

In order to visualise the spectral variability of the source, we extracted the \textit{Swift}/XRT light curve of NGC 5204 X-1 in the $0.3-10$ keV energy band, using the website tool\,\footnote{https://www.swift.ac.uk/user\_objects/} \citep{evans2009}. The \textit{Swift}/XRT monitoring was performed from April 2013 to February 2025.
As the XMM-\textit{Newton} spectra from all observations show flux variability below approximately $3$ keV (see also \citealt{gurpide2021a}), we computed the spectral hardness of the \textit{Swift}/XRT data as the ratio between the flux in the $3-10$ keV and in the $0.3-3$ keV energy bands. 
Figure \ref{fig:xrt_lightcurve} shows the \textit{Swift}/XRT light curve and hardness ratio, with the dashed lines marking the start times of the two most recent deep XMM-\textit{Newton} observations, as well as the histogram of the count rates derived from the light curve.

The observational data files (ODF) were downloaded from the \xmm Science Archive (XSA) \footnote{https://www.cosmos.esa.int/web/\xmm/xsa} and we performed the standard data reduction procedure using the \textit{Science Analysis System} ({\scriptsize{SAS}}) version 21.0.0\,\footnote{https://www.cosmos.esa.int/web/\xmm} with recent calibration files (April 2024). 
We used data from the European Photon Imaging Camera (EPIC) and the Reflection Grating Spectrometers (RGS) onboard XMM-\textit{Newton}. The EPIC cameras are useful to provide an accurate description of the spectral continuum, while RGS is used to detect narrow spectral features in the soft X-ray band. 
The cleaned events files of the EPIC-pn and MOS 1,2 cameras were created through the \textit{epproc} and \textit{emproc} tasks, respectively. 
Periods of high background rate (e.g. due to Solar flares) were excluded using the \textit{evselect} task by considering the light curves above 10 keV, with thresholds of 0.5 c/s for pn and 0.35 c/s for MOS 1,2. In addition, we filtered the data to account for bad pixels and bad columns. Events with PATTERN<=4 (single/double) were accepted for pn data, while for MOS 1, 2 data, events with PATTERN<=12 (up to quadruple) were considered.

To extract spectra, we chose a circular region for the source of $20$ arcsecond radius centred at the source position (\textit{Chandra} coordinates: RA = 13$^{\mathrm{h}}$ 29$^{\mathrm{m}}$ 39$^{\mathrm{s}}$, Dec = +58$^{\circ}$ 25\arcmin\,06\arcsec). For the background, a circular region of 50 arcsecond radius was selected away from the source on the same chip, in order to avoid the contamination from the copper ring and the chip gaps.
To produce response matrices and effective area files, we used the \textit{rmfgen} and \textit{arfgen} tasks.
We also produced stacked spectra for each EPIC camera using \textit{epicspeccombine} to deepen our search for features in the full time-averaged spectra obtained by combining all observations, and in the archival data (hereafter "archive-only"), to validate the result of \citet{Kosec2018a} and corroborate our new method.

The events files of RGS 1,2 were created using the \textit{rgsproc} task, which also produces spectra and response matrices. Periods of high background rate were filtered out by examining the background light curve from RGS 1,2 CCD number 9 and rejecting time bins with a count rate exceeding 0.2 c/s. 
Subsequently, we extracted the 1$^{\text{st}}$-order spectra by considering 90\% of the source point-spread-function, and the background spectra by selecting photons beyond 98\% of the point-spread-function. The individual RGS 1 and 2 spectra from observations 0921360101 and 0921360201 contain about 2200 net source counts each. As the individual 1$^{\text{st}}$-order RGS 1 and 2 spectra of each observation were superimposable in the band of interest ($0.45-1.77$\,keV), we stacked them using \textit{rgscombine} to decrease the number of spectra to be analysed and speed up the computational time, which is necessary for Monte Carlo simulations aimed at estimating the significance of line detection (see Sect.\,\ref{sec:MC_simulations}).

As previously done for EPIC cameras, we also stacked the RGS spectra for the archive-only data obtaining a spectrum fully compatible with that of \citet{Kosec2018a} and a very deep time-averaged spectrum corresponding to a total exposure of about 420 ks and roughly 11000 net source counts. 
We also extracted the $2^{\mathrm{nd}}$-order RGS 1 and 2 spectra, but since they are highly affected by the instrumental background, we did not use them in the analysis.

The spectral analysis was performed using \texttt{SPEX} version 3.07.03 \citep{spex,kaastra2022}. 

\section{Spectral analysis} \label{sec:analysis}
In this section, we present the spectral analysis of NGC 5204 X-1. We first focused on modelling the spectral continuum. Next, we performed a Gaussian line scan to search for and identify any spectral features. Finally, we tested different physical plasma models to study the nature of the spectral features and the structure of the outflows responsible for them.
The spectra were grouped according to the optimal binning criterion as defined in \citealt{kaastra2016} using the \texttt{obin} command in \texttt{SPEX}, and were fitted by minimising the Cash statistics (\textit{C}-stat; \citealt{cash}). The errors in the model parameters correspond to $1\sigma$ confidence level.
We adopted Solar abundances from \citet{lodders2009}, which are default in \texttt{SPEX}.
The spectral ranges used for our spectral analysis are: $0.3-0.5$ keV and $1.77-10$ keV for EPIC, and $0.45-1.77$ keV (i.e. $7-27$ \AA{}) for RGS because for lower energies the background dominates the source spectrum. We ignored the EPIC data in the $0.5-1.77$ keV range to employ the high spectral resolution of RGS and reduce the degeneracy between emission and absorption line models, while still maintaining a good handle on the continuum through the EPIC data (e.g. \citealt{Pinto2021}). When including EPIC data in the $0.5-1.77$ keV range, the continuum parameters vary at the $10-20\%$ level, consistent with continuum uncertainties.
We notice that in this work we do not perform a dedicated broadband, time-resolved study, in particular, of the evolution of the continuum properties which will be the focus of a companion paper (Barra et al., in prep.).

\subsection{Continuum modelling}
\begin{figure*}
   \centering
   \includegraphics[width=\textwidth]{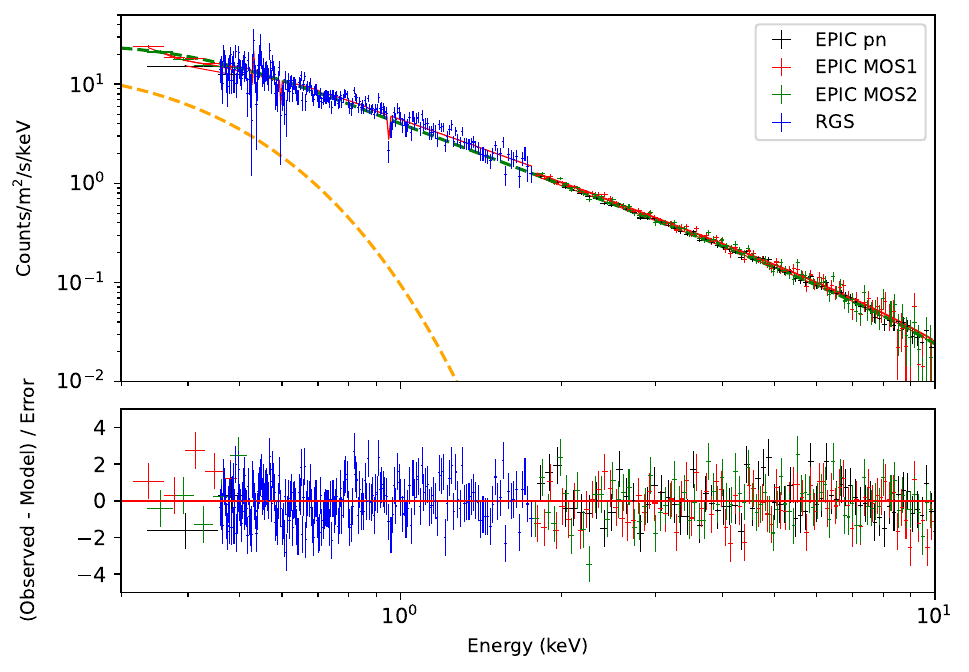}
   \vspace{-0.6cm}
   \caption{EPIC-pn (black), MOS 1 (red), MOS 2 (green) and RGS (blue) spectra and the best-fit continuum model (red curve), from the stacking of all XMM-\textit{Newton} observations. The continuum emission components are represented by dashed lines in orange (\texttt{bb}) and green (\texttt{comt}).
   The EPIC data are ignored in the $0.5-1.77$ keV energy range to employ the full high spectral resolution of RGS and avoid model degeneracy. The EPIC data are grouped according to the optimal binning and, for a better visualisation, the RGS data are grouped by a factor of 9. The bottom panel shows the residuals of the best-fit model.}
    \label{fig:continuum_spectra}
\end{figure*}

We simultaneously fitted the EPIC-pn, MOS 1,2 and RGS full time-averaged spectra. To account for cross-instrument calibration uncertainties, which we found to be between $2\%$ and $10\%$, we included a free multiplicative constant in the model, fixing it to unity for EPIC-pn. 
The spectral continuum is modelled as \texttt{hot}*(\texttt{bb}+\texttt{comt}), comprising two emission components. 
The \texttt{bb} component represents the blackbody emission from the outer parts of the accretion disc or the wind, while the \texttt{comt} component describes the Comptonisation of soft photons in a hot plasma. This process might originate from the scattering of photons emitted within the geometrically thick disc (see e.g. \citealt{Middleton2015a}). Additionally, we coupled the temperature of the \texttt{bb} component to the temperature of the seed photons of the \texttt{comt} component.
The absorption by the foreground interstellar and circumstellar media was accounted for using the \texttt{hot} component, and we fixed the temperature at $10^{-6}$ keV, to describe a neutral gas.
We also tested other continuum models such as \texttt{hot}*(\texttt{dbb}+\texttt{comt}), where \texttt{dbb} describes the emission of a multi-temperature disc blackbody, but the fit remained approximately unchanged.
The best-fit hydrogen column density, $N_\text{H}$, is $5.4^{+0.6}_{-0.8}\times 10^{20}$ cm$^{-2}$, about twice the Galactic value estimated by \citealt{dickey1990} and implies additional absorption in the NGC 5204 galaxy. The time-averaged spectra, along with the best-fit continuum model and residuals, are shown in Fig.\ref{fig:continuum_spectra}. It is possible to observe some residuals around 0.6 keV and 1 keV, which hint for the presence of spectral lines.

A spectral fitting for the characterisation of the continuum is also performed for the combined archive-only data and for the two most recent observations, to reproduce the results already obtained by \citet{Kosec2018a} and to study the variability of the spectra between observations in two different flux regimes. In this and all subsequent analyses, the value of $N_\mathrm{H}$ was fixed to that derived from the fit of the time-averaged data, as the amount of plasma in the interstellar medium is not expected to vary significantly across observations.
The best-fit results for the different datasets are shown in Table \ref{table:continuum}.

\begin{table*}
\begin{center}
\caption{Results of the best-fit continuum modelling.}  
\label{table:continuum}     
\renewcommand{\arraystretch}{1.2}
 \small\addtolength{\tabcolsep}{4pt}
 \vspace{-0.1cm}
\scalebox{0.95}{%
\hskip-0.0cm\begin{tabular}{@{}ccccc}     
\hline  
Parameter  &  Archive-only & All-times & 0921360101 & 0921360201  \\
\hline
\multicolumn{5}{c}{Model: \texttt{hot}*(\texttt{bb}+\texttt{comt})} \\
\hline
 L$_{\text{X,bb}}$ ($10^{39}$ erg/s) & $0.35\pm0.04$ & $0.32^{+0.03}_{-0.04}$ & $1.0 \pm 0.2$ & $<0.16$ \\
 L$_{\text{X,comt}}$ ($10^{39}$ erg/s) & $4.9\pm0.3$  &  $4.9^{+0.2}_{-0.3}$ & $4.0\pm0.3$ & $5.9 \pm 0.6$ \\
 kT$_{\text{bb}}$ (keV) & $0.100^{+0.001\,(a)}_{-0}$ & $0.100^{+0.005\, (a)}_{-0}$ & $0.14 \pm 0.01$ & $0.12 \pm 0.01$ \\
 kT$_{\text{in, comt}}$ (keV) & $0.1^{(b)}$  & $0.1^{(b)}$  & $0.14^{(b)}$  & $0.12^{(b)}$ \\
 kT$_{\text{e, comt}}$ (keV) & 2.44$^{+0.19}_{-0.15}$ & 2.68$^{+0.18}_{-0.15}$  &  $2.33^{+0.18}_{-0.14}$ & $3.4^{+2.0}_{-0.7}$ \\
 $\tau_{\text{comt}}$ & $5.8\pm 0.3$ & $5.4\pm0.2$  & $6.6^{+0.3}_{-0.4}$  & $4.0^{+0.7}_{-1.3}$ \\
 \textit{C}-stat/d.o.f. & 850/714 & 903/756  & 942/769  & 1002/846 \\
\hline                          
\end{tabular}
}

\vspace{0.2cm}

Notes: The X-ray luminosities of the plasma emission components are computed in the $0.3-10$ keV energy range, assuming a distance of $4.5$ Mpc. $^{(a)}$ indicates that the parameter is pegged at the lower boundary of the allowed range. $^{(b)}$ indicates coupled parameters: for each dataset, $\mathrm{kT}_{\text{bb}}= \mathrm{kT}_{\text{in, comt}}$. The hydrogen column density, $N_{\text{H}}$, of the interstellar medium is fixed at $5.4\times10^{20} \text{cm}^{-2}$, as obtained for the time-averaged data.
\vspace{-0.3cm}
\end{center}
\end{table*}

\subsection{Gaussian line scan}
\begin{figure*}
   \centering
   \includegraphics[scale=0.55]{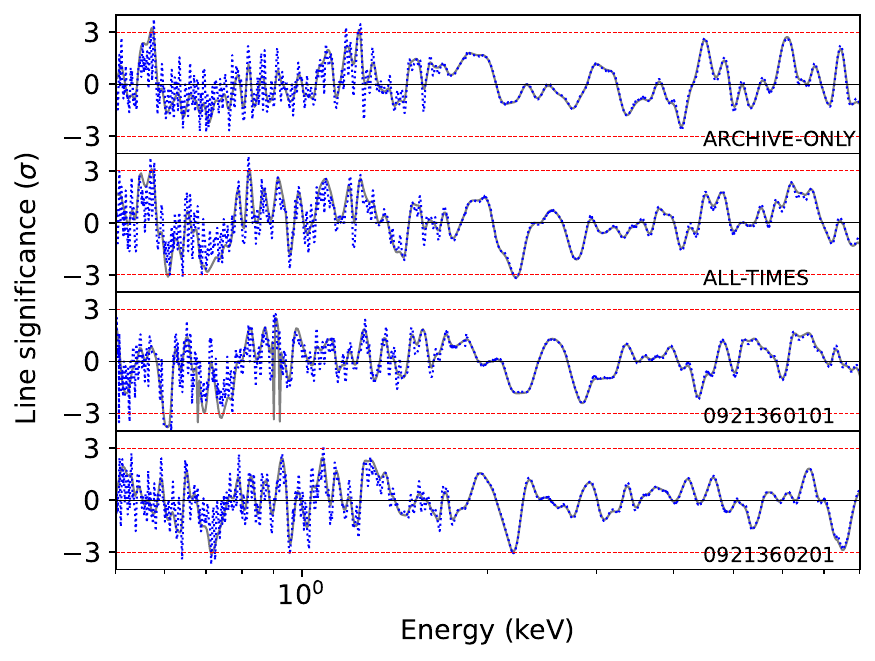}
   \hspace{0.6cm}
   \includegraphics[scale=0.55]{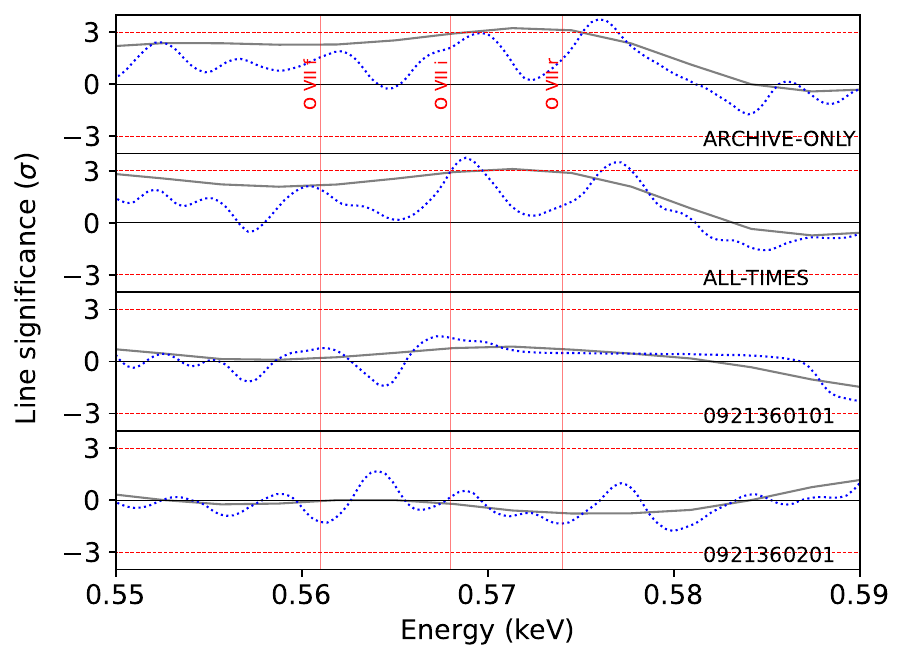}
   \vspace{-0.3cm}
   \caption{Gaussian line scan performed on the four analysed datasets. We show the results for velocity widths of 2500 km/s (solid grey line) and 100 km/s (dotted blue line). The $3\sigma$ threshold (red line) is shown. In the right panel, a zoom in the $0.5-0.6$ keV energy range is shown to better visualise the O VII line triplet, with the red lines showing the positions of the rest-frame energies.}
    \label{fig:gaussian_scan}
   \vspace{-0.2cm}
\end{figure*}

In order to search for spectral lines due to the presence of outflows, we first performed a scan of the spectra with a moving Gaussian line. 
We added a Gaussian line to the best-fit continuum model using the \texttt{gaus} component in \texttt{SPEX} with a specific centroid energy and line width. 
The model was fitted for the normalisation of the line allowing for both positive and negative values, corresponding to emission and absorption lines, respectively, along with the continuum. The \textit{C}-stat was saved and compared to the value obtained by the fit of the continuum-only model, computing the $\Delta C$-stat improvement.
We then varied the centroid energy of the line over a logarithmic grid ranging from 0.5 to 8 keV, searching for spectral lines also outside the RGS energy range -- where the EPIC source spectra lie well above the background. The number of grid points ranged from 500 to 4000, as we varied the velocity dispersion, $\sigma_v$, from 2500 to 100 km/s. 
The single-trial significance of each line was calculated as the square root of the $\Delta C$-stat improvement times the sign of the normalisation of the line to distinguish between absorption and emission features.
Nevertheless, this method does not take into account the look-elsewhere effect, e.g. spurious features associated with Poisson noise, resulting in an overestimate of the significance of individual lines \citep{protassov2002}. The real significance of the spectral features can be estimated by running Monte Carlo simulations (see Section \ref{sec:MC_simulations}).
The results of the line scan are shown in Figure \ref{fig:gaussian_scan}.

We first performed the line search in the combined archive-only data. There are strong emission features with a $\Delta C$-stat $\sim 12$ at 1.2 keV ($\sim 10$ \AA) and at about 0.57 keV ($\sim 22$ \AA), analogously to what is found in \citealt{Kosec2018a}. 
Then, we performed the line search on the full time-averaged spectra and on the individual observations.
At about 2.2 keV, for the time-averaged spectra and for the observation 0921360201, we see an absorption feature, due to an absorption edge characteristic of the instrument, likely corresponding to the Au-M edge.
For the time-averaged spectra, we find strong emission features at about 0.8 keV ($\sim15$ \AA) with a $\Delta C$-stat $\sim 14$ and at about 0.57 keV ($\sim 22$ \AA) with a $\Delta C$-stat $\sim 12$. There is evidence of an absorption feature at 0.6 keV ($\sim20$ \AA), also found in the observation 0921360101 ($\Delta C$-stat $\sim17$). In the observation 0921360201, there is evidence of an absorption feature at about 0.7 keV ($\Delta C$-stat $\sim14$). 
Specifically, the emission lines found in common between the archive-only and time-averaged data at about $\sim 22$ \AA\,correspond to the O VII triplet, with the resonance line (\textit{r}) at 21.6 \AA, intercombination line (\textit{i}) at 21.8 \AA, and forbidden line (\textit{f}) at 22.1 $\AA$ (see Figure \ref{fig:gaussian_scan}). The resonance line seems to be blueshifted, hinting at an outflow moving towards us.
We modelled the O VII triplet in the time-averaged data with simple Gaussian profiles to obtain some information about the physical state of the plasma by calculating the \textit{G}-\textit{R} line ratios, as illustrated in Section \ref{sec:ovii}.
There is no significant evidence of the O VII triplet or other strong emission lines in the two most recent observations.

\subsection{Physical model scans}
In order to identify the nature of the spectral features, we employed physical models of line emission or absorption, performing spectral scans in a multidimensional parameter space for a plasma in collisional-ionisation equilibrium (CIE) and photoionisation equilibrium (PIE). This approach allows us to simultaneously fit multiple lines reducing the risk of getting stuck in local minima and allowing for the identification of multiple phases of the plasma. We searched through the parameter space of the line-of-sight velocity, $v_{\text{LOS}}$, and the key plasma property (the temperature, kT, for CIE or the ionisation parameter, $\xi$, for PIE). This method essentially builds on the previous works of \citet{Kosec2018b} and \citet{Pinto2020b}.

\subsubsection{Collisionally-ionised plasma in emission} \label{sec:CIE}

\begin{figure*}
   \centering
   \includegraphics[scale=0.45]{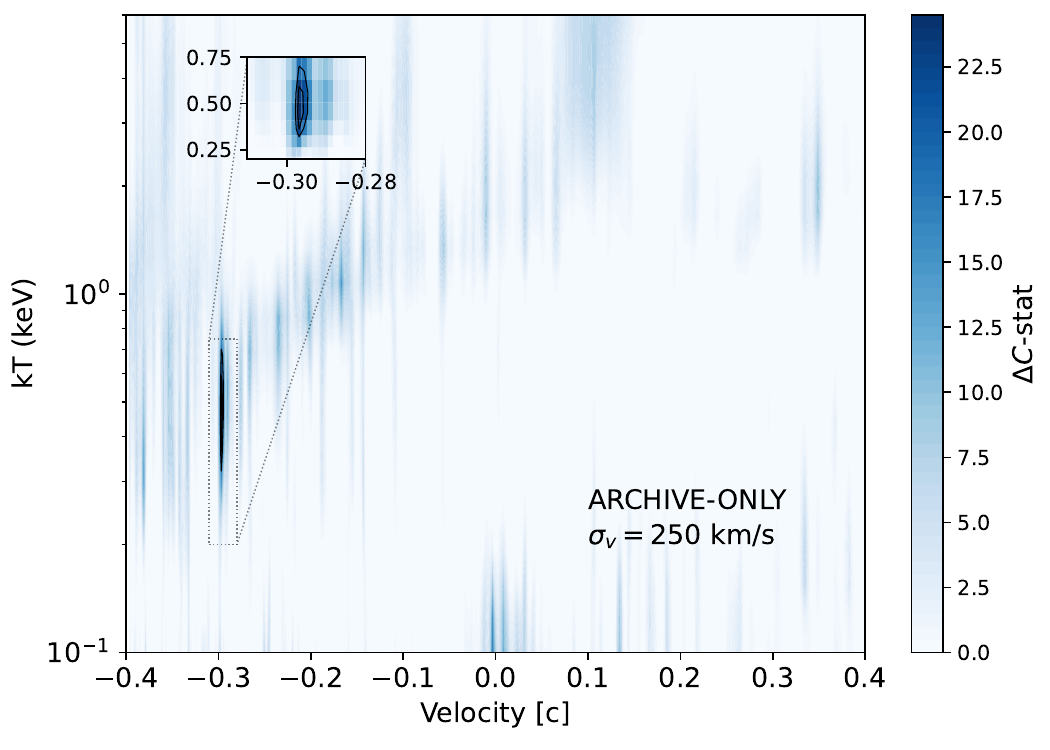}
   \hspace{0.5cm}
   \includegraphics[scale=0.45]{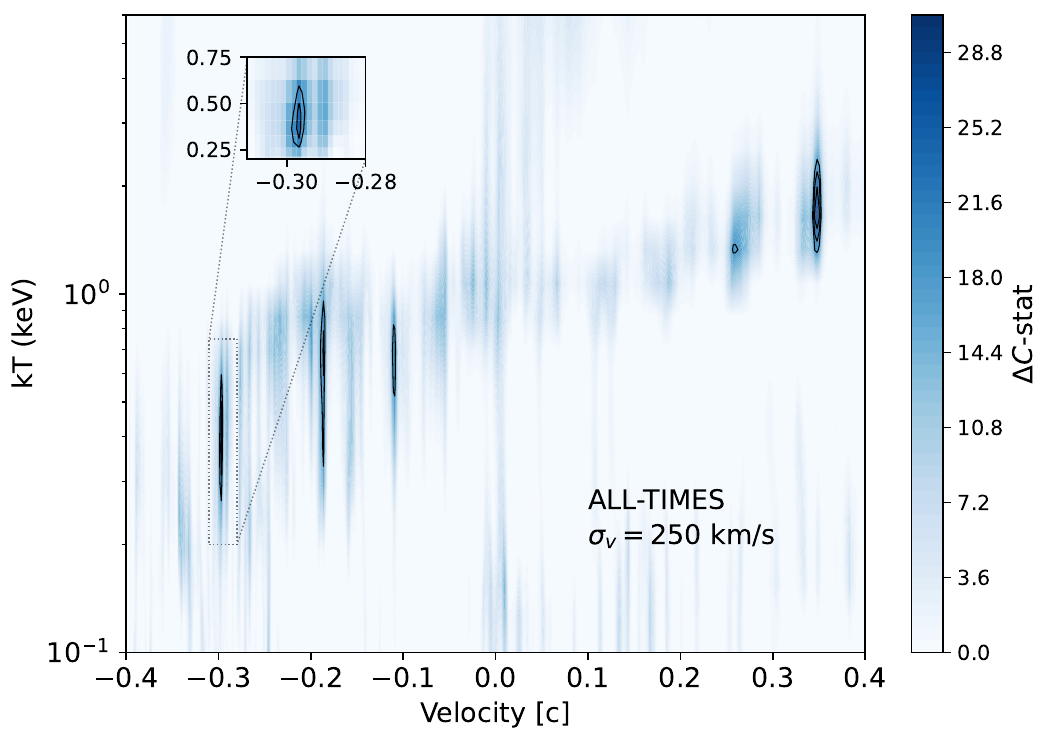}
   \includegraphics[scale=0.45]{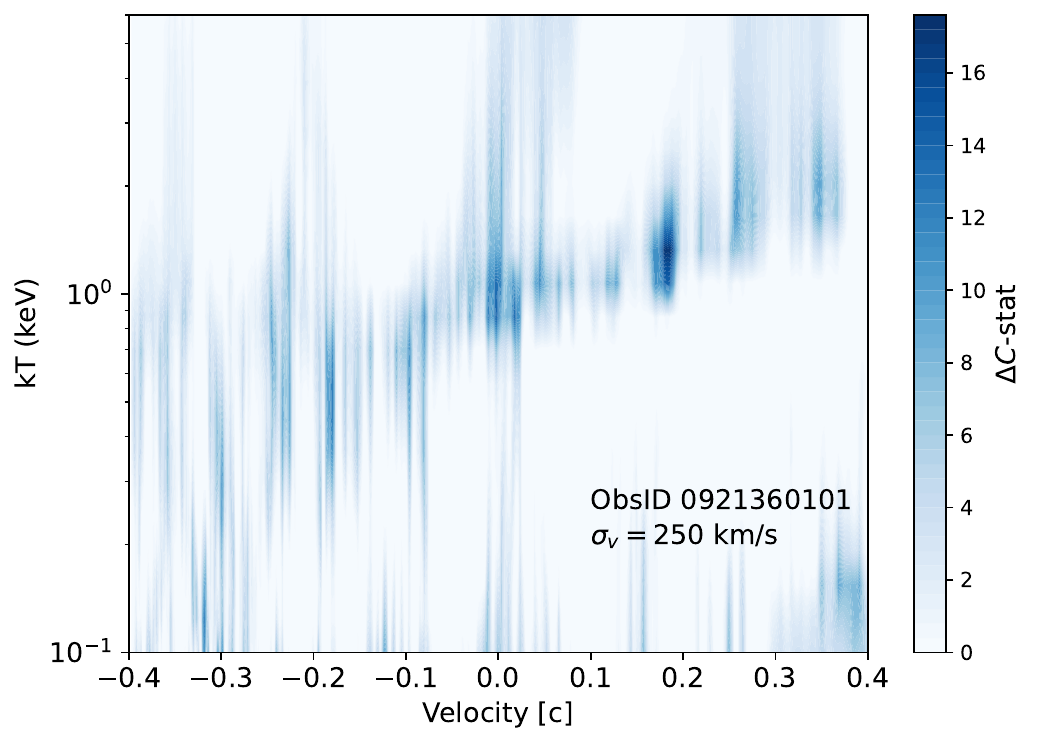}
   \hspace{0.5cm}
   \includegraphics[scale=0.45]{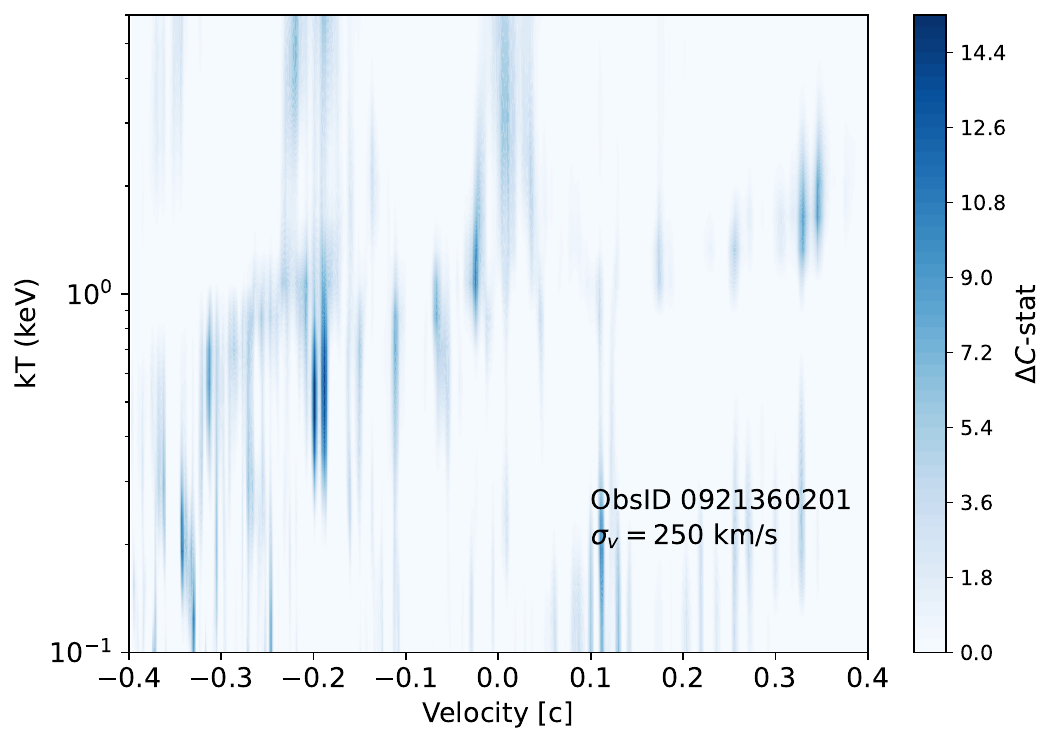}
   \vspace{-0.3cm}
   \caption{Collisional-ionisation equilibrium model scan performed on the four analysed datasets, adopting a velocity dispersion of $250$ km/s. The colours show indication of the $\Delta C$-stat improvement compared to the continuum-only model. 
   The black contours refer to significance levels from 2.5 to 3.5 sigma with steps of $0.5\sigma$ estimated with Monte Carlo simulations (see Section \ref{sec:MC_simulations}). 
   }
    \label{fig:cie_scans}
\end{figure*}

In the case of a line-emitting CIE plasma, we adopted the \texttt{cie} model in \texttt{SPEX}, which calculates the emitting spectrum for a plasma in collisional-ionisation equilibrium. Additionally, to account for possible blueshifted and/or redshifted emission lines, we added the \texttt{reds} component, which multiplies the \texttt{cie} model, with FLAG=1 to apply a Doppler shift rather than a cosmological redshift.
We constructed a grid of plasma temperatures, kT, line-of-sight velocities, $v_{\mathrm{LOS}}$, and velocity dispersions, $\sigma_v$.
We adopted a logarithmic grid of temperatures spanning the range $0.1-6$ keV, sampled at 20 points, and a linear grid of line-of-sight velocities from $-0.4c$ (blueshifted outflow) to $+0.4c$ (redshifted outflow). The velocity dispersion was decreased from 2500 to 100 km/s, with steps from 1000 to 500 km/s for the line-of-sight velocity.
The normalisation of the \texttt{cie} component corresponding to the emission measure, $n_{\text{H}}n_eV$ - with $n_e$ and $n_{\text{H}}$ denoting the electron and hydrogen densities, respectively, and \textit{V} the volume of the line-emitting plasma - was left free to vary.
For each combination of (kT, $v_{\text{LOS}}$, $\sigma_v$), we recorded the $\Delta C$-stat improvement with respect to the continuum-only model. In Fig. \ref{fig:cie_scans}, we show the results of the scan for the four different datasets. The CIE solutions are found mainly at kT $\lesssim$ 1.5 keV, which means that the results are driven primarily by the RGS data.

Our analysis of the combined archive-only data finds a fit improvement with respect to the continuum model, with a $\Delta C$-stat $= 24$ for a velocity dispersion of $250$ km/s, a \texttt{cie} temperature of approximately $0.45$ keV and a line-of-sight velocity of $-0.3c$ (see Table\,\ref{table:physical_models}). \citet{Kosec2018a} find a solution that corresponds to a secondary peak in our maps at $-0.33c$.
Additionally, we also find an independent solution with a $\Delta C$-stat = 16 which describes a lower-temperature plasma (kT$\sim0.1$ keV) close to rest-frame. This component is primarily responsible for the O VII line emission.
For other values of the velocity dispersion, we do not obtain larger fit improvements.

From the analysis of the time-averaged spectra, we find two line-emitting components: one with a temperature of about $0.4$ keV and outflow velocity of approximately $-0.30c$ ($\Delta C$-stat $=24$, see Table\,\ref{table:physical_models}), another component with a temperature of $1.65$ keV and an outflow velocity of about $0.35c$ ($\Delta C$-stat $=30$). Another solution is found at a temperature of approximately $0.2$ keV, which describes an approximately rest-frame plasma ($\Delta C$-stat $=18$).
Additional solutions appear at velocities of about $-0.2c$ with temperatures comparable to those of the components discussed above -- potentially reproducing similar spectral features -- but with lower $\Delta C$-stat values. The technical evaluation of secondary solutions is performed in Appendix\,\ref{sec:appendix_secondary}.

For observations 0921360101 and 0921360201, the fit improvement is lower ($\Delta C$-stat $=15-17$).
%Furthermore, as the temperatures are approximately the same, we expect to detect in our spectra the same spectral features at different energies due to the different Doppler shift.

For each dataset, we then fitted the EPIC and RGS spectra adding to the continuum the \texttt{cie} models with higher $\Delta C$-stat values to better describe the spectra and identify the contribution of each \texttt{cie} to the observed emission lines (see Section \ref{models_fit}). 
 
\subsubsection{Photoionised plasma}\label{sec:photoionisation}
A plasma in photoionisation equilibrium is described by the ionisation parameter, $\xi=L_{ion}/{n_{\text{H}}R^2}$, where $L_{ion}$ is the luminosity of the ionising source, $n_{\text{H}}$ is the hydrogen number density and \textit{R} is the distance of the plasma from the source.
The ionisation balance depends on the relative elements abundance; in this work, we adopted Solar abundances as for the interstellar medium absorber.
We computed the ionisation balance using the \texttt{pion} code (\citealt{Mehdipour2016,miller2015}), which performs an instantaneous calculation of the balance, adopting the best-fit continuum model (see Table \ref{table:continuum}) as the spectral energy distribution (SED). 
Another way of computing the ionisation balance consists of the use of the \texttt{xabsinput} tool in \texttt{SPEX}, which produces tables of the temperature and ionic column density as functions of the ionisation parameter, $\xi$. These outputs are used by the \texttt{xabs} model to compute the spectrum of a photoionised plasma in absorption. We obtain similar results with both methods.

\begin{figure}
   \centering
   \includegraphics[scale=0.5]{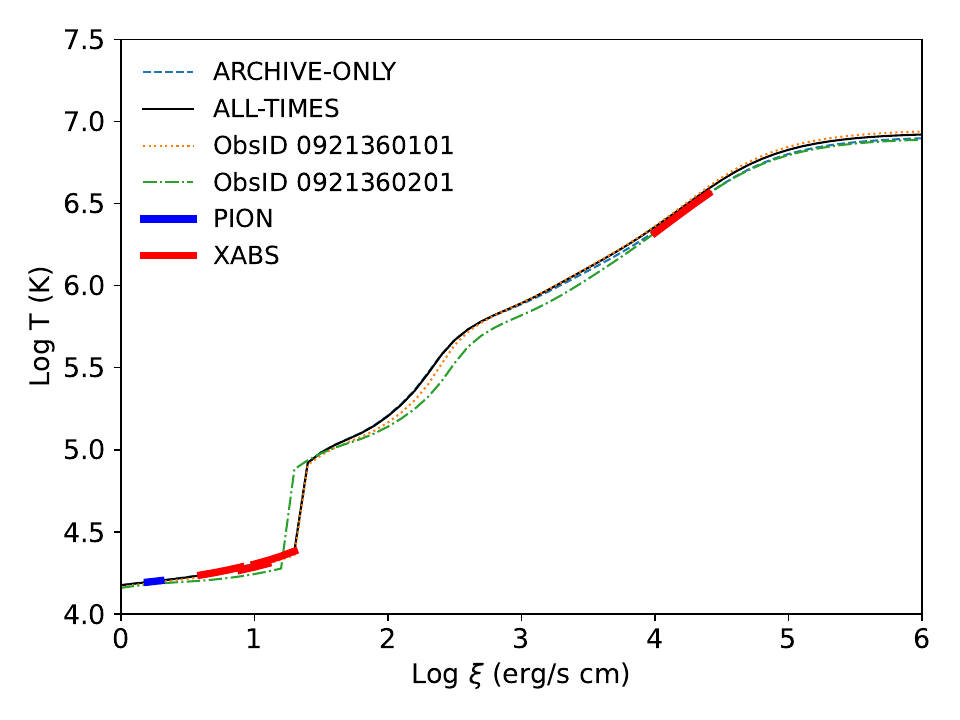}
   \includegraphics[scale=0.5]{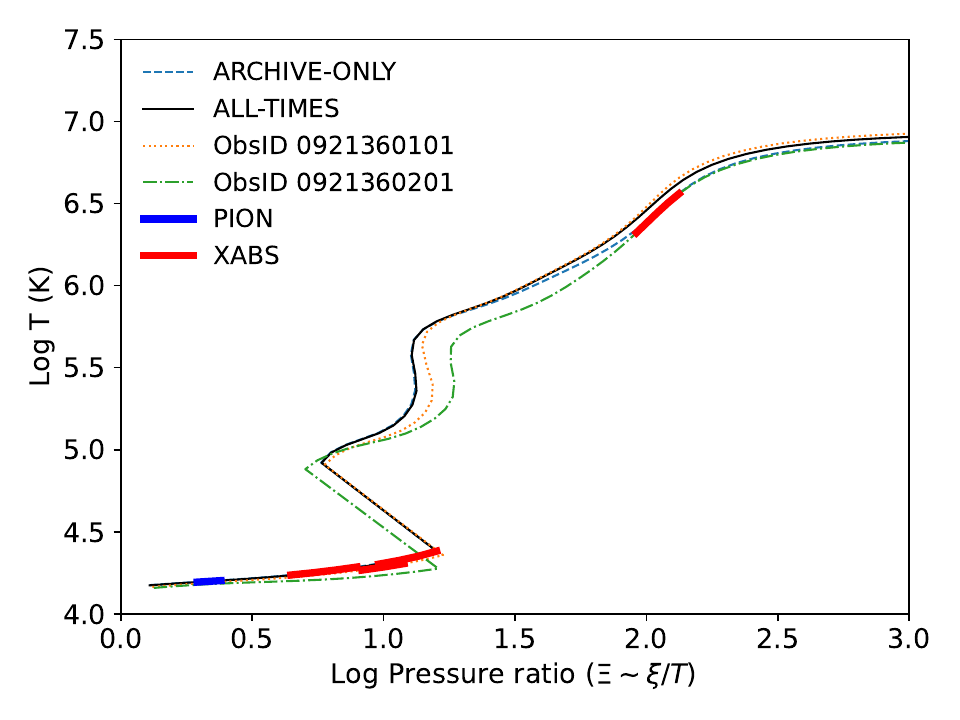}
   \vspace{-0.4cm}
   \caption{Ionisation balance (top) and stability curve (bottom) for the time-averaged spectra, archive-only data, and new individual observations, using as the SED the best-fit continuum model for each case. Thick blue and red segments show the ranges of the best-fit solutions from the \texttt{pion} and \texttt{xabs} models.}
    \label{fig:sed_ionbal}
   \vspace{-0.2cm}
\end{figure}

The ionisation balance depends on the relationship between the ionisation parameter and the temperature, which is illustrated in Figure \ref{fig:sed_ionbal}. An alternative way for illustrating the ionisation bal-
ance is through the stability curve, also known as \textit{S} curve, which shows the temperature as function of the ratio between the radiation pressure ($L/4\pi R^2 c$) and the thermal pressure ($n_H$kT), which is given by $\Xi=19222\, \xi/$T. Along this curve heating equals cooling, i.e. the gas is in thermal balance \citep{Krolik1981}. 
The photoionised plasma is thermally stable for $\Xi$ values corresponding to positive slopes of the \textit{S} curve, while negative slopes indicate thermal instability to temperature perturbations. The presence of unstable branches may produce a multiphase plasma. We would not typically expect to find (T$,\Xi$) solutions occupying unstable branches unless short-lived events occur.

Line-emitting plasma. In the case of the emitting plasma in photoionisation equilibrium, we adopted the \texttt{pion} component, which was included additively in our spectral model, on top of the continuum, and a \texttt{reds} component was applied to account for Doppler shifts. Additionally, we adopted for \texttt{pion} a full solid angle of $\Omega=4\pi$, representing the emission from a thin shell, and a covering fraction, $f_{\text{cov}}$, of zero, to produce only emission lines.
We performed a spectral scan using the \texttt{pion} model creating a grid of ($v_{\text{LOS}}$, $\xi$, $\sigma_v$). Specifically, we varied $\log\xi$ from 0 to 6 with 0.1 steps, and scanned $v_{\text{LOS}}$ from $-0.4c$ to $+0.4c$ to reproduce both blueshifted and redshifted emission lines. We tested different velocity dispersions ($\sigma_v$= 100, 250, 500, 1000, 2500 km/s).
The column density, $N_{\text{H}}$, of the \texttt{pion} component was left free to vary.
The results of the scans for the four datasets are shown in Fig. \ref{fig:pion_scans}. 

%We observed that in the \texttt{pion} model scans of the four different datasets, the $\Delta C$-stat is lower compared to the case where a \texttt{cie} model is adopted.
Specifically, from the analysis of the time-averaged spectra, we obtain a fit improvement of $\Delta C$-stat $=24$ for a blueshift of about 3900 km/s and an ionisation parameter of $\log \xi \sim 0.2$ (corresponding to a T $\sim1.5\times10^4$ K), assuming a velocity dispersion of $\sigma_v=250$ km/s. A smaller improvement ($\Delta C$-stat $= 17$) but with similar values of $v_\mathrm{LOS}$ and $\log \xi$ is obtained from the scan of the archive-only data. 
Additionally, the scan of the time-averaged spectra reveals another improvement ($\Delta C$-stat $=20$) for $v_\mathrm{LOS}\sim-0.2c$ and $\log \xi \sim 2.7$ (corresponding to a T $\sim6\times10^5$ K), under the same assumption of $\sigma_v=250$ km/s. Similarly, the archive-only data shows a fit improvement of $\Delta C$-stat $=21$ for a $v_\mathrm{LOS}\sim-0.05c$ and $\log \xi \sim 0.1$.
The scan of individual observations show similar but less significant results, with no remarkable improvements in observation 0921360201.

%101: soluzione a -0.013c con v_sigma=100 deltaC-stat=18, soluzione a -0.2c con v_sigma=250 deltaC-stat=17

Line-absorbing plasma. In the case of the absorbing plasma in photoionisation equilibrium, we computed the ionisation balance through the \texttt{xabsinput} tool in \texttt{SPEX}. We then performed a spectral scan using the faster \texttt{xabs} model exploring the $\xi-v_{\mathrm{LOS}}$ parameter space. The value of $\log\xi$ was varied from 0 to 6 with 0.1 steps, and $v_{\text{LOS}}$ from $-0.4c$ to 0, to reproduce only blueshifted absorption lines due to outflows. We tested different velocity dispersions ($\sigma_v$ = 100, 250, 500, 1000, 2500 km/s).
For each combination of ($\log\xi$, $v_{\text{LOS}}$, $\sigma_{v}$), we fitted the column density, $N_\mathrm{H}$, of the \texttt{xabs} component. In Fig. \ref{fig:xabs_scans}, we show the results of the scans for the different datasets.

For ObsID 0921360101, higher values of $\Delta C-$stat are found; specifically, we obtain a fit improvement of $\Delta C$-stat = 23 for a $v_\mathrm{LOS}\sim-0.05c$ and $\log \xi \sim 1$, adopting $\sigma_v=2500$ km/s.
The same is observed for the archive-only and time-averaged data, but with a much lower $\Delta C$-stat ($< 13$).
For ObsID 0921360201, we obtain a fit improvement of $\Delta C$-stat $=17$ for a $v_\mathrm{LOS}\sim-0.09c$ and $\log \xi\sim4.2$, adopting $\sigma_v=1000$ km/s.

\subsection{Significance estimates through Monte Carlo simulations} \label{sec:MC_simulations}

The significance of the detection of a line-emitting or line-absorbing plasma through the use of physical model grids is obtained by performing identical multi-parameter searches for a large number of spectra simulated with \texttt{SPEX}, which only contain Poisson noise (and no outflow features) and are based on the best-fit continuum model. This enables us to account for the look-elsewhere effect. We used the method outlined by \citet{Pinto2020b}.

\begin{figure}
   \centering
  \includegraphics[scale=0.5]{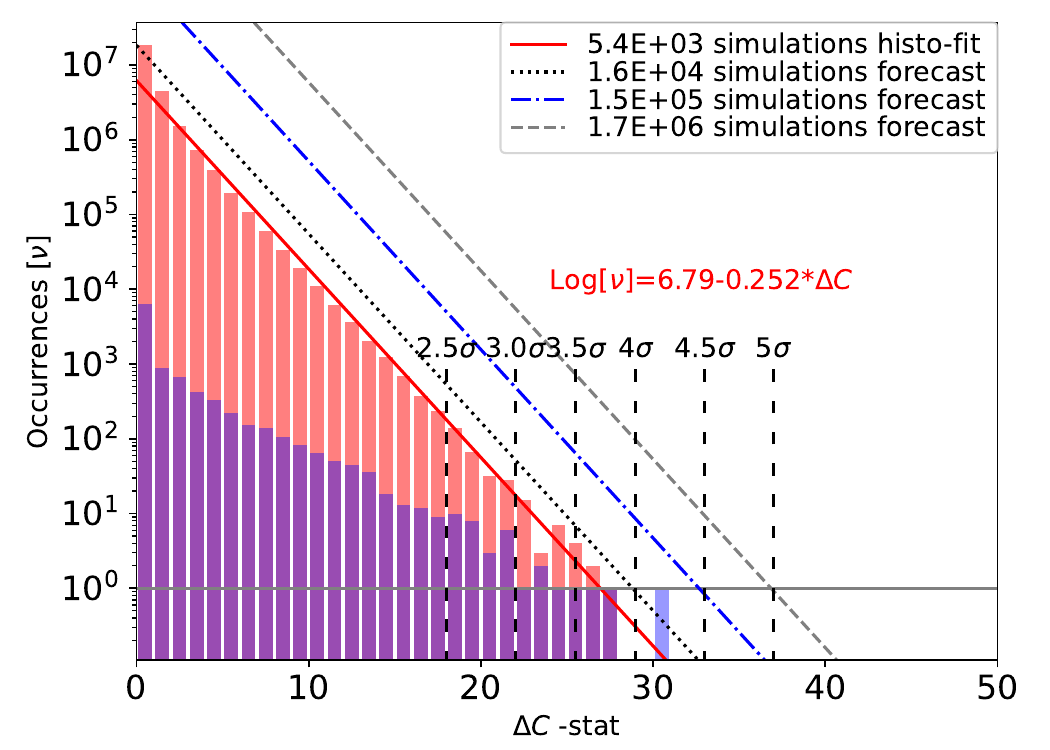}
   \vspace{-0.3cm}
   \caption{Histogram of $\Delta C$-stat values of a CIE model scan of the time-averaged data adopting a velocity dispersion of $250$ km/s (blue bars), and of CIE model scans of 5,400 Monte Carlo simulated data (red bars). Forecasts for more simulations are made assuming a constant slope of the histogram, in order to visualise different confidence levels.}
    \label{fig:simulations}
   \vspace{-0.3cm}
\end{figure}

In the case of NGC 5204 X-1, we simulated 5,400 EPIC and RGS spectra, adopting as a template the best-fit continuum model of the EPIC+RGS observation with ObsID 0921360101, and with the same exposure times of the real data.
Each simulated spectrum was scanned with the \texttt{cie} model, using the same parameter space explored for the real data (Section \ref{sec:CIE}).
The histogram of the $\Delta C$-stat values was produced to estimate the significance for the detection in the real data. This is shown in Figure \ref{fig:simulations}. We adopted a velocity dispersion of ${\sigma}_v=1000$ km/s, which is midway between the best-fit results of the individual components (see Table \ref{table:physical_models}). The parameter space for this velocity dispersion is slightly smaller than that for narrower values (e.g. 250 km/s, for which a velocity step of 500 km/s was used instead of 700 km/s). However, this difference is not significant, as it remains well within the RGS spectral resolution. Indeed, a set of a thousand simulations performed with the narrower step yields a negligible change in the significance ($\sim0.1\sigma$), which is within the uncertainty of the histogram slope (see Figure \ref{fig:simulations}).

Since the histogram slope ($\sim -0.252$) is approximately constant after a few thousand simulations, it is possible to forecast the results for a larger number of simulations re-scaling the normalisation of the best-fit power law (see \citealt{Pinto2021}, \citealt{gu2022}).
The false-alarm probability is estimated computing the fraction of simulations with the same or higher $\Delta C$-stat than the value obtained for the real data.
According to the simulations, $\Delta C$-stat $>22\,(29)$ would correspond to a confidence level above $3\sigma\, (4\sigma)$. The high significance of the observed spectral features allows us to make a comparison between the data and complex models, avoiding the risk of over-interpreting the data.

We repeated the same procedure for the time-averaged data for a thousand simulations, and the histogram of the $\Delta C$-stat values has a comparable slope.

\subsection{Final fits with physical models} \label{models_fit}

In order to investigate the nature of the plasma responsible for the observed emission/absorption lines, we performed spectral fits for the archive-only, full time-averaged, and the two most recent individual spectra, considering the results obtained by the physical model scans.
Analogously to the analysis performed in \cite{Kosec2018a}, we analysed the combined archive-only data by adding a blueshifted \texttt{cie} model on top of the continuum.  
The improvement of the fit yields a $\Delta C$-stat $=26$ for four degrees of freedom ($n_\mathrm{H}n_eV$, kT, $\sigma_v$, $v_\mathrm{LOS}$), adopting a volume density of $10^{10} \mathrm{cm}^{-3}$, estimated from the analysis of the O VII triplet (see Section \ref{sec:ovii}). In particular, we assume that the volume density of the relativistic outflow is identical to that of the slow-moving plasma, as the available emission lines do not allow us to constrain or distinguish the densities of the two plasma components.
The \texttt{cie} component exhibits a blueshift of $(-0.2960\pm0.0006)c$ and a temperature of  $0.49^{+0.07}_{-0.08}$ keV. We obtain a $1\sigma$ upper limit on the velocity broadening of 500 km/s. The emission features of the \texttt{cie} component are reproduced by Fe XVII and O VIII.
Additionally, a further fit improvement ($\Delta C$-stat = 15 respect with the previous model) is achieved with the addition of a rest-frame \texttt{cie} component, along with the blueshifted one. The plasma temperature of the rest-frame component is $0.12 \pm 0.02$ keV and the velocity dispersion is $2000 ^{+3000}_{-1000}$ km/s.
This second \texttt{cie} component reproduces the O VII emission, with a dominant resonance line as expected for a hybrid plasma where collisions are not negligible. The centroid of the resonance line seems to be blueshifted, while the intercombination and forbidden lines appear to be at rest. 

\begin{figure*}
   \centering
\includegraphics[width=0.9\textwidth]{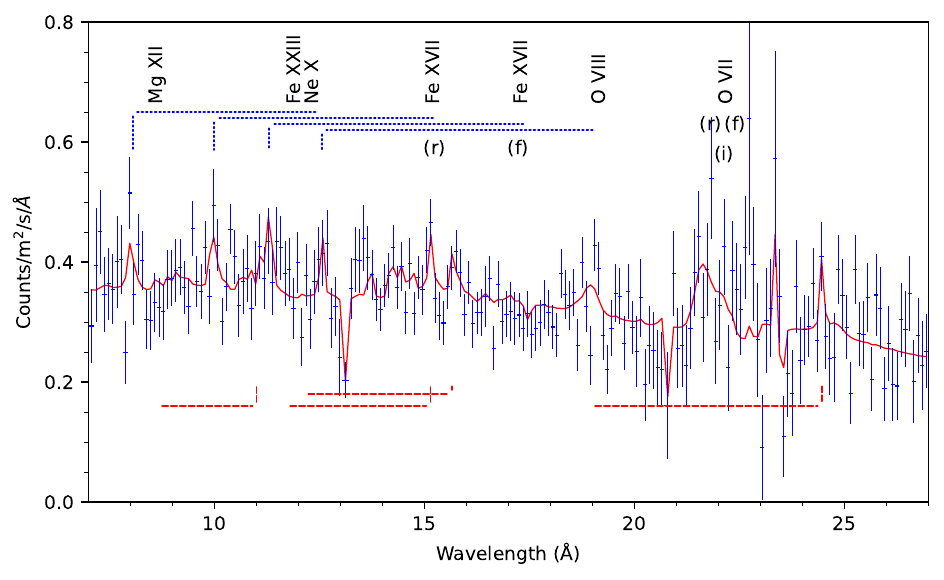}
   \vspace{-0.4cm}
   \caption{RGS spectrum in the $7-27$ \AA{} range, obtained by stacking all XMM-\textit{Newton} observations. The red curve shows the model with three \texttt{cie} components (blueshifted, redshifted and rest-frame), on top of the continuum. The rest-frame wavelengths of the most relevant lines are labelled. The dotted (dashed) lines show the velocity shift for the blueshifted (redshifted) lines. The data are grouped by a factor of 10.}
    \label{fig:cie_bestfit_model}
\end{figure*}

\begin{figure}
   \centering
   \includegraphics[scale=0.5]{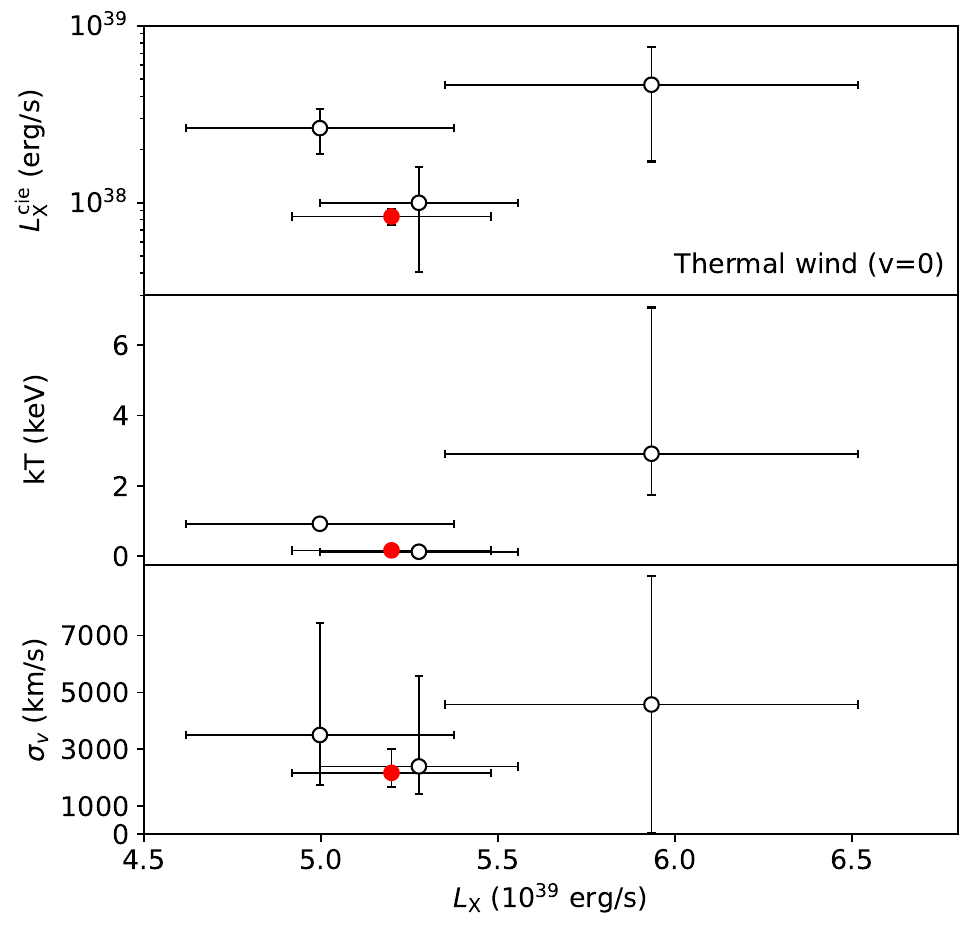}
    %\vspace{-0.2cm}
   \includegraphics[scale=0.5]{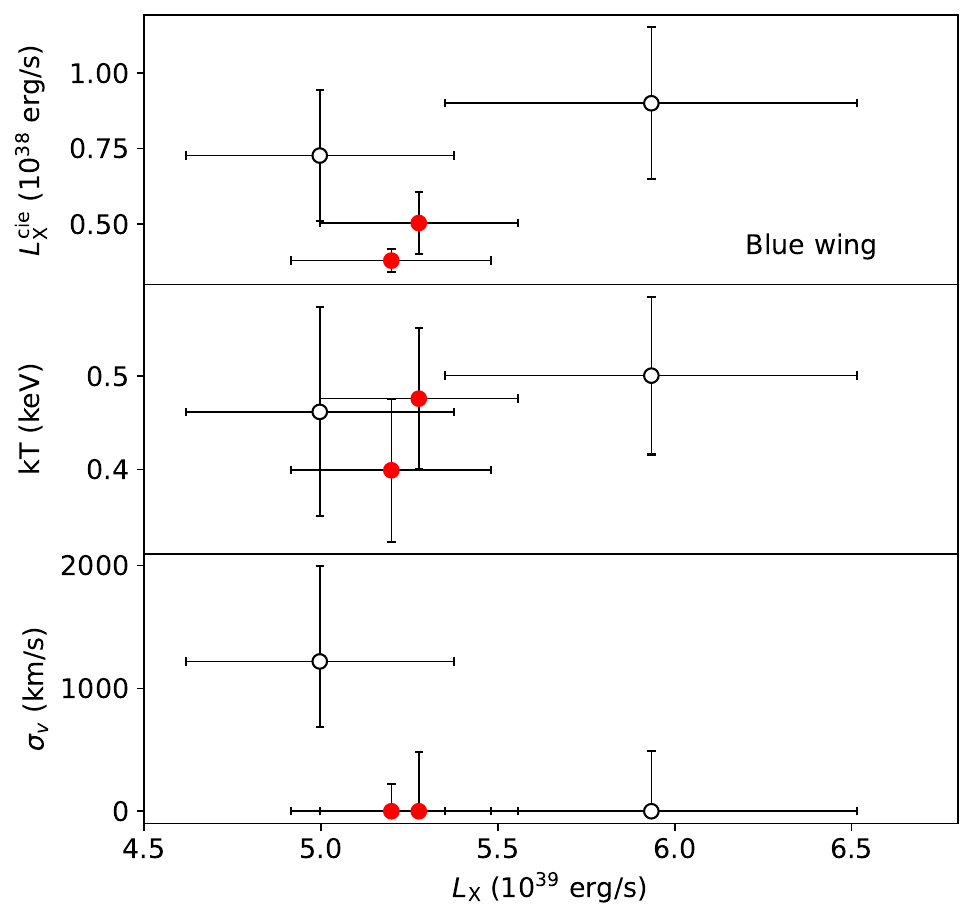}
   %\vspace{-0.2cm}
   \includegraphics[scale=0.5]{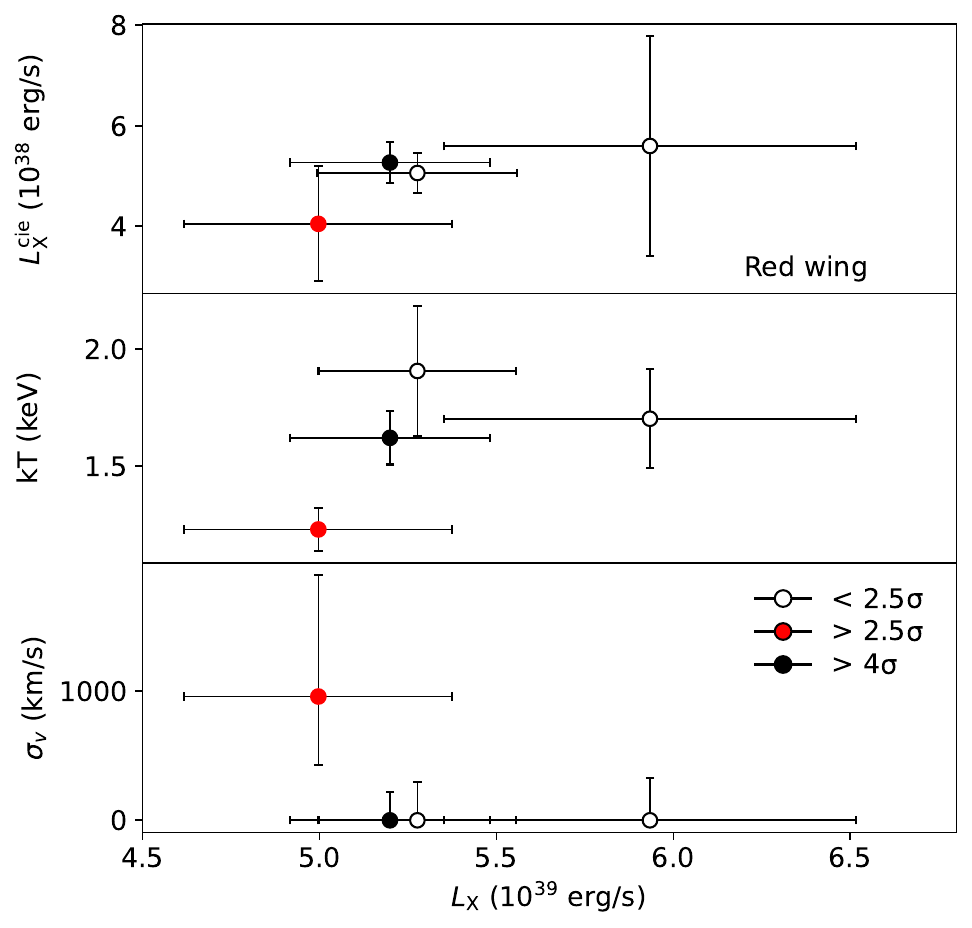}
   \vspace{-0.1cm}
   \caption{The X-ray ($0.3-10$ keV) luminosity, temperature, and velocity dispersion of the \texttt{cie} plasma components as functions of the continuum X-ray luminosity. Red filled (empty) dots show solutions with significance above (below) $2.5\sigma$, and black filled dots are those above $4\sigma$.} %%% The X-ray luminosities are computed in the $0.3-10$ keV energy range.
    \label{fig:CIE_results}
\end{figure}

For the time-averaged spectra, based on the results of the physical model scans, we added two \texttt{cie} components to the continuum model. A blueshift was applied to the first and a redshift to the second and the velocity dispersion was linked between the two components. A total improvement of $\Delta C$-stat $=57$ compared to the continuum-only model is achieved ($\Delta C$-stat $=25$ and $32$ for the individual blueshifted and redshifted \texttt{cie} components, respectively), confirming that the two solutions are independent and account for distinct spectral features. 
The first \texttt{cie} component exhibits a blueshift of $(-0.2966^{+0.0010}_{-0.0002})c$ and a temperature of $0.40^{+0.10}_{-0.05}$ keV, consistent with the results from the archive-only data, while the second \texttt{cie} component shows a redshift of $(0.3473^{+0.0005}_{-0.0007})c$ and a temperature of $1.62^{+0.13}_{-0.10}$ keV, with an $1\sigma$ upper limit on the velocity broadening of 200 km/s. The blueshifted \texttt{cie} model fits several emission lines, reproduced by Ne X, Fe XVII and O VIII. Specifically, the dominant lines are the resonance ($\lambda=15$~\AA) and forbidden ($\lambda=17.1$~\AA) lines of Fe XVII, observed at $\sim 10$~\AA{} and $\sim 11.3$~\AA{}, respectively.
The redshifted component fits other emission features, reproduced by Ne X, Fe XXIII, Fe XXIV and O VIII.
%The addition of a third \texttt{cie} component with a blueshift of $-0.19c$ improves the fit with a further $\Delta$C-stat = 13, which means that the solution is not fully independent and that some degeneracy might take place.
None of these fast components is able to reproduce the O VII line triplet, which is close to the laboratory wavelengths and was already observed in the archive-only data.
Consequently, we tested the addition of a rest-frame \texttt{cie} component ($\Delta C$-stat = 18) and a weakly blueshifted \texttt{pion} component ($\Delta C$-stat = 22).
The \texttt{pion} component exhibits a blueshift of $4200^{+180}_{-120}$ km/s and a value of $\log \xi=0.20^{+0.11}_{-0.06}$. We also fitted the volume density, obtaining a value of $1.343^{+0.017}_{-0.008}\times10^{12}$ cm$^{-3}$, consistent with the results from the analysis of the O VII triplet (see Section \ref{sec:ovii}). 
This model reproduces the O VII emission, where the forbidden line is stronger than the other lines, as expected from a photoionised plasma.
Figures \ref{fig:cie_bestfit_model} and \ref{fig:pion_allexp} show the best-fit models with the rest-frame \texttt{cie} and \texttt{pion} components, respectively. Similar analyses were performed for the two recent individual observations, with all best-fit results reported in Table \ref{table:physical_models} and Figure \ref{fig:CIE_results}.

%In both of them, we found evidence of plasma outflowing at velocities of approximately $-0.3c$ and $0.3c$, although with low $\Delta C$-stat improvements ($\sim 10 - 12$). For this reason, we focused on fitting other, more statistically significant features identified in the physical model scan.
%For the observation 0921360101, we added two \texttt{cie} components, applying a redshift to the first one. The fit yielded a $\Delta C$-stat $\sim 35$; the first \texttt{cie} component exhibits a line-of-sight velocity of $(0.180\pm0.004)c$ and a temperature of $1.20^{+0.09}_{-0.08}$ keV, while the second \texttt{cie} component shows a temperature of $1.5^{+0.3}_{-0.2}$ keV.
%For observation 0921360201,  we added a \texttt{cie} component. The best fit yields a $\Delta C$-stat $\sim 12$, representing a plasma outflowing at $-0.194^{+0.010}_{-0.005}c$ with a temperature of $0.5^{+0.3}_{-0.1}$ keV.}
Furthermore, as suggested from the scan of PIE plasma in absorption for observation 0921360101, the addition of a \texttt{xabs} component improves the fit, obtaining $\log\xi=1.00 \pm 0.12$, $v_\mathrm{LOS}=(-0.051\pm0.004)c$, a $\Delta C$-stat of about 23 for a velocity dispersion of $1900^{+1700}_{-600}$ km/s. Using this model, the absorption feature we observed at 0.6 keV would be attributed to O VII.

\begin{table*}
\begin{center}
\caption{Best-fit parameters of the \texttt{cie} models applied on the continuum model for the four analysed datasets. \texttt{cie}$_0$, \texttt{cie}$_\mathrm{B}$ and \texttt{cie}$_\mathrm{R}$ refer to the rest-frame, blueshifted and redshifted \texttt{cie} components, respectively.
The X-ray luminosities of the plasma emitting components are computed in the $0.3-10$ keV energy range. $^{(a)}$ indicates fixed parameters.
%The $\Delta C$-stat$^\mathrm{a}$($^\mathrm{e}$) indicates the improvement in the fit statistics when the component is included alone in the model (a) or together with the other two components (e), in addition to the continuum.
}
\label{table:physical_models}     
\renewcommand{\arraystretch}{1.06}
 \small\addtolength{\tabcolsep}{4pt}
 \vspace{-0.1cm}
\scalebox{0.95}{%
\hskip-0.0cm\begin{tabular}{@{}cccccc}     
\hline 
Model & Parameter  &  Archive-only & All-times & 0921360101 & 0921360201  \\
\hline
 \makecell{\texttt{cie}$_0$ \\ (Thermal wind)} & $v_\mathrm{LOS}(c)$  & $=0^{(a)}$ & $=0^{(a)}$ &  $=0^{(a)}$ &  $=0^{(a)}$ \\
 & $\sigma_v$ (km/s)   & $2000^{+3000}_{-1000}$ & $2000^{+800}_{-500}$  & $4000^{+4000}_{-2000}$  &  $5000\pm5000$ \\
& kT (keV) & $0.12\pm0.02$ & $0.16 \pm 0.02$ & $0.92\pm0.07$ &  $2.9^{+4.1}_{-1.2}$ \\
& $\mathrm{L}_\mathrm{X}^{\mathrm{cie}}\, (10^{38} \mathrm{erg/s})$ &  $1.0\pm0.6$ &                $0.84\pm0.08$ & $2.6\pm0.8$ & $5\pm3$\\
& $\Delta C$-stat & 15 & 18 & 17 & 6 \\
\hline   
 \makecell{\texttt{cie}$_\mathrm{B}$ \\ (Blue wing)} & $v_\mathrm{LOS}(c)$  & $-0.2960\pm0.0006$ &   $-0.2966^{+0.0010}_{-0.0002}$ &  $-0.181^{+0.002}_{-0.107}$ &  $-0.1987\pm0.0008$  \\
 & $\sigma_v$ (km/s)   & $<500$ & $<200$ & $1000^{+800}_{-500}$  &  $<500$                               \\
& kT$^\mathrm{cie}$ (keV) & $0.48 \pm 0.08$ & $0.40^{+0.10}_{-0.05}$ & $0.50\pm0.08$ &  $0.50\pm0.08$ \\
& $\mathrm{L}_\mathrm{X}^{\mathrm{cie}}\, (10^{38} \mathrm{erg/s})$ &  $0.5\pm0.1$ &                 $0.38\pm0.04$ & $0.7\pm0.2$ & $0.9\pm0.3$\\
& $\Delta C$-stat & 26 & 25 & 14 & 15 \\
\hline   
 \makecell{\texttt{cie}$_\mathrm{R}$ \\ (Red wing)} & $v_\mathrm{LOS}(c)$  & $0.348\pm0.001$ &   $0.3473^{+0.0005}_{-0.0007}$ &  $0.182^{+0.003}_{-0.004}$ &  $0.345\pm0.001$  \\
 & $\sigma_v$ (km/s)   & $<300$ & $<200$ &  $900^{+900}_{-500}$  &  $<300$ \\
& kT$^\mathrm{cie}$ (keV) & $1.9^{+0.3}_{-0.2}$ & $1.62^{+0.13}_{-0.10}$ & $1.2\pm0.1$ &  $1.70^{+0.30}_{-0.13}$\\
& $\mathrm{L}_\mathrm{X}^{\mathrm{cie}}\, (10^{38} \mathrm{erg/s})$ & $5\pm2$  &                 $5.3\pm0.4$ & $4.0\pm1.1$ & $6^{+3}_{-2}$\\
& $\Delta C$-stat & 10 & 32 & 18 & 10 \\
\hline
\end{tabular}
}
\vspace{-0.5cm}
\end{center}
\end{table*}

\section{Discussion} \label{sec:discussion}
\subsection{Physical properties of the line-emitting plasma} \label{sec:ovii}

\begin{figure}
   \centering
   \includegraphics[scale=0.5]{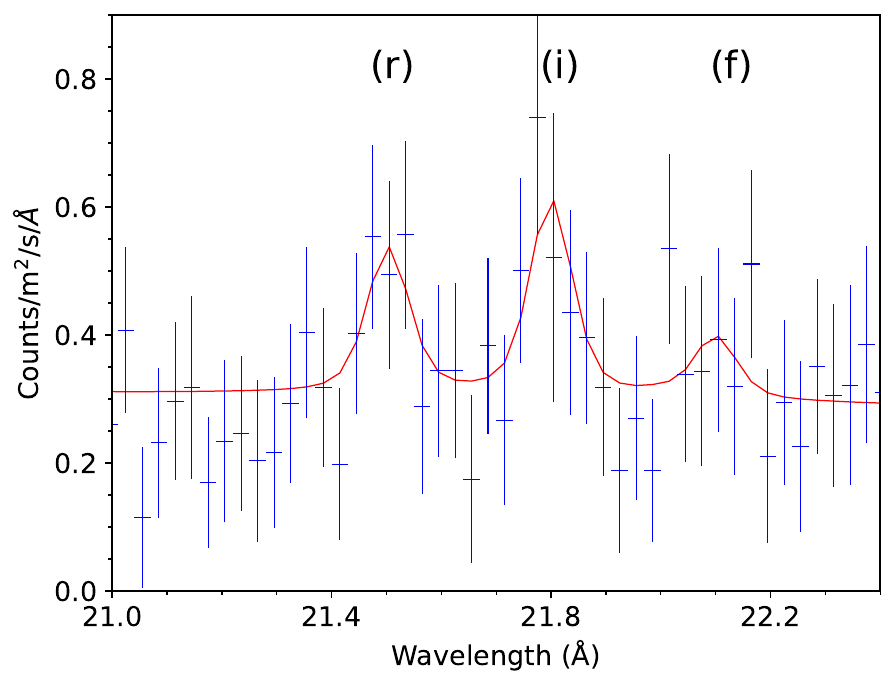}
   \vspace{-0.3cm}
   \caption{Zoom on the O VII emission triplet (r: resonance, i: intercombination, f: forbidden; see Section \ref{sec:ovii}) as fitted with two sets of three Gaussian lines (red) in the RGS time-averaged spectrum (blue).}
    \label{fig:zoom_ovii}
   \vspace{-0.3cm}
\end{figure}

As first diagnostics of the plasma properties, we constrain the electronic density and temperature through the calculation of the \textit{R} and \textit{G} ratios for the He-like triplets, which are defined as follows \citep{gr_ratios}: $R(n_e)=\textit{f}/\textit{i}$ and $G(T_e)=(\textit{f}+\textit{i})/\textit{r}$, where \textit{r}, \textit{i} and $\textit{f}$ are the fluxes of the resonance, intercombination and forbidden lines, respectively.
We computed the \textit{R} and \textit{G} ratios for the O VII triplet, for which we found evidence in the Gaussian line scan (see Figure \ref{fig:gaussian_scan}), by modelling the emission lines as simple Gaussian lines in the time-averaged spectra.
As shown in Figure \ref{fig:gaussian_scan}, a single set of three Gaussian lines does not provide a satisfactory fit, due to clear evidence of a blueshifted O VII resonance line. We therefore added two sets of three O VII lines to the best-fit continuum model: one at rest-frame energies, and a second set with a common blueshift applied, with all line energies fixed to their laboratory values.
Additionally, their FWHMs were coupled. We fitted the line normalisations along with the continuum (see Figure \ref{fig:zoom_ovii}). The best-fit yields a $\Delta C$-stat $=26$ compared to the continuum-only model. For the rest-frame O VII triplet, we obtain $R=0.4\pm0.3$ and a lower limit of $G>8$, the latter resulting from the upper-limit on the normalisation of the resonance line. For the blueshifted emission lines, we find $R\sim1.56$, with only upper limits on the normalisations of the intercombination and forbidden lines, and an upper limit of $G<0.44$. The blueshifted set of O VII lines have a common line-of-sight velocity of about $1500 \pm 300$ km/s.

\begin{figure*}
   \centering
   \includegraphics[scale=0.35]{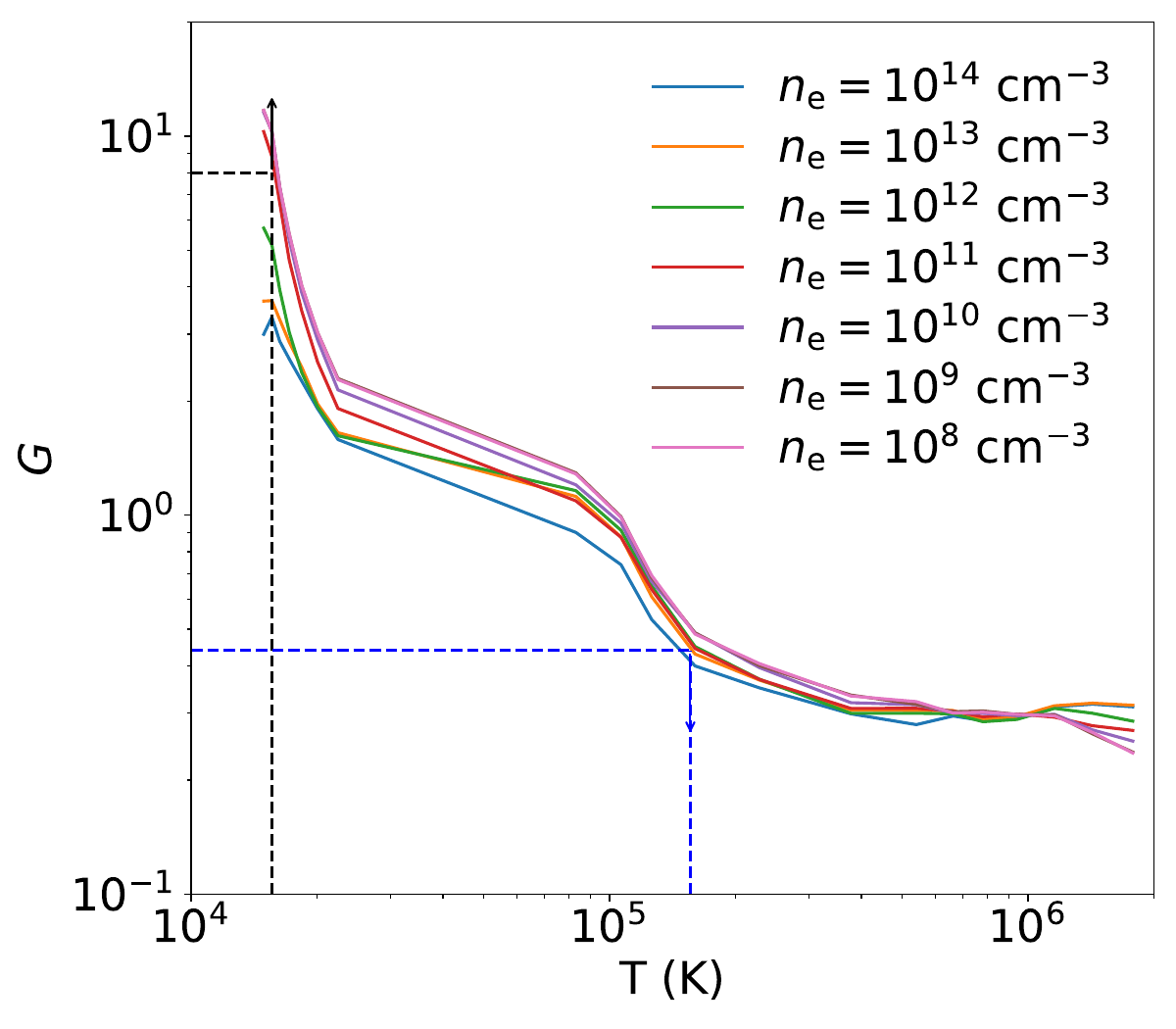}
   \hspace{0.6cm}
   \includegraphics[scale=0.35]{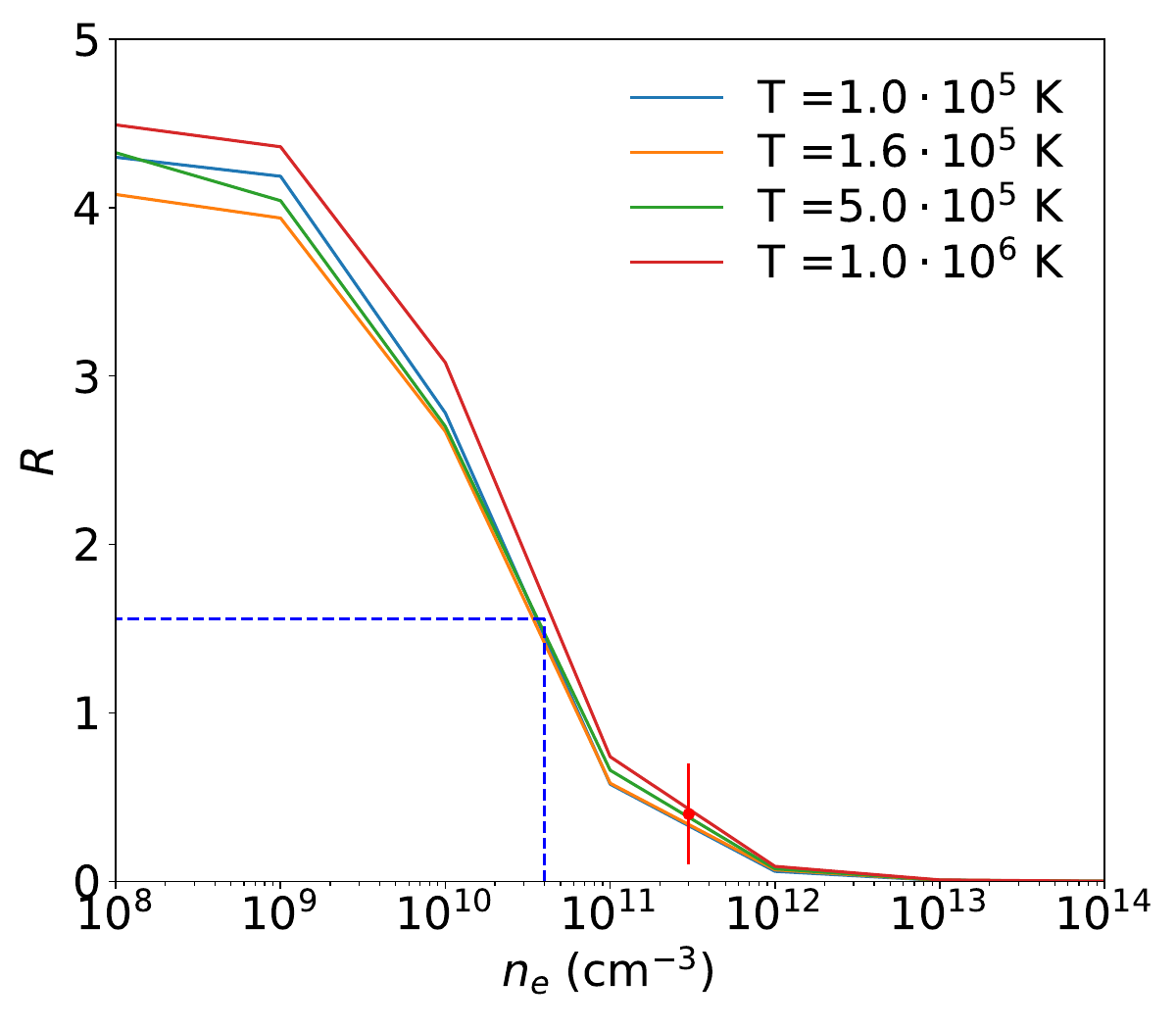}
   \vspace{-0.3cm}
    \caption{\textit{G} (left) and \textit{R} (right) ratios curves for the O VII triplet. Blue arrows show upper and lower limits, corresponding to lower and upper limits on the temperature, respectively. The black dashed lines indicates the results obtained for the rest-frame O VII triplet, while the blue dashed lines for the blueshifted emission lines. In the right panel, the red point indicates the value of \textit{R} obtained from the rest-frame O VII lines.}
    \label{fig:ovii_line_ratios}
   \vspace{-0.3cm}
\end{figure*}

We computed the \textit{R} and \textit{G} ratios for the O VII triplet as functions of the electronic density, $n_e$, and the electronic temperature, $T_e$, respectively, using the \texttt{SPEX} \texttt{pion} code.
The best-fit continuum model of the time-averaged spectra was used as the SED; the photoionisation balance was obtained with the \texttt{ascdump} command, along with the fluxes of the O VII lines.
The computation was realised by varying $\log{\xi}$ between 0 and 4.0 with 0.2 steps, and $n_e$ from $10^{8}$ to $10^{14}\,\text{cm}^{-3}$ with a logarithmic step of 10. We also adopted a velocity dispersion of $250$ km/s, as it provides the best-fit improvement. The \textit{G} and \textit{R} ratios curves are shown in Fig. \ref{fig:ovii_line_ratios}.  
Using the values of \textit{R} and \textit{G} obtained by the fitting of the rest-frame O VII emission lines, we obtain a number density around $3 \times 10^{11} \text{cm}^{-3}$, with an upper limit of $10^{12} \mathrm{cm}^{-3}$, and $T_{e} \leq 1.5\times10^{4}$ K, indicative of a PIE plasma. Based on the values of \textit{R} and \textit{G} from the blueshifted emission lines, we infer a number density around $4 \times 10^{10} \text{cm}^{-3}$ and a temperature $T_e \geq 1.5\times10^{5}$ K, suggesting both a heating process and a CIE plasma.
The values found for $n_e$ are consistent with those found for the ULX NGC 1313 X-1 \citep{1313x1_pi}.

\subsection{Outflow structure and constraints on the ULX nature}

The spectral scan of the time-averaged spectra with a collisionally-ionised plasma model revealed evidence for two plasma components moving in opposite directions along the line of sight, both with velocities of about $0.3c$. Consequently, the two most recent deep observations helped us to investigate the structure of the outflow, revealing its biconical geometry.
The archive-only data analysed in \citet{Kosec2018a} only allowed the detection of the collisionally-ionised plasma blueshifted to about $0.3c$.
However, the individual observations do not provide sufficient statistics to identify significative variations in the outflow properties. 
As shown in Figure \ref{fig:CIE_results}, the outflow parameters are consistent across the different datasets, indicating the presence and stability of the outflow, as also supported by the higher significance obtained from the full time-averaged data.

Specifically, for the component outflowing at approximately $-0.3c$, we obtain a $\Delta C$-stat $=25$, resulting in a significance $>3\sigma$, while for the component moving at about $0.3c$, the $\Delta C$-stat is $32$, corresponding to $>4\sigma$. The significances were estimated through Monte Carlo simulations. The luminosity of the first component is $4\times10^{37}$ erg/s, while that of the second component is an order of magnitude higher ($5\times10^{38}$ erg/s). This is attributed to the blueshifted plasma being found to be cooler than the redshifted counterpart, possibly due to interactions with regions of the interstellar medium having different physical properties. Nevertheless, both the luminosities and the corresponding temperatures are consistent within a few $\sigma$ (see Figure \ref{fig:cie_scans}).
%Based on Doppler boosting, we would expect the approaching component to appear brighter.

Our results are similar to those reported for the Galactic source SS433, as previously noted in \citet{Kosec2018a}, where the emission originates from oppositely directed plasma jets in collisional-ionisation equilibrium. The jets are accelerated to $0.27c$ and, accounting for projection effects, the blue jet is blueshifted by $0.08c$ and the red jet is redshifted by $0.16c$ \citep{ss433_lines}. Some lines are in common to our time-averaged spectra and the spectra of SS433, such as the Ne X Ly$\alpha$, which is observed at $11.9$ \AA{} for the blue jet of SS433 and at $8$ \AA{} for NGC 5204 X-1. Additionally, our source shows the Mg XII emission line in the redshifted component of the outflow, which is also present in the blue jet of SS433.
Although we find some similarities, we have to consider their different viewing angle, given that SS433 is observed edge-on, while NGC 5204 X-1 is seen at a lower inclination.

Assuming that the observed outflows in NGC 5204 X-1 correspond to perfectly opposed jets, using Equations (2) and (3) from \citet{marshall2013}, a jet velocity of $\beta\sim0.3$ yields an angle between the jets and the line of sight of $\lesssim 50^{\circ}$. Consequently, the line of sight does not intersect the densest regions of the fast outflow, reducing strong absorption while allowing emission from the sides of the outflow. In this context, the collisionally ionised nature of the fast components may originate from shock heating in regions that do not directly intercept the brightest continuum source.

Additionally, a precession of the outflows similar to that observed in SS433 may be present, which would produce secular or quasi-periodic variations in the Doppler shifts and in the relative strength of the two components.
However, the signal-to-noise ratio of the individual observations is insufficient to robustly track the Doppler shifts as a function of time, as exposures longer than 100 ks would be required. The available data therefore do not allow us to place meaningful constraints on possible velocity variations. The two latest observations show weak detections of moving plasma components with different dominant contributions at about $-0.2c$ and $0.2c$, indicating a possible variation in the relative strengths of the blue- and redshifted components. In addition, by scanning the time-averaged spectrum using the \texttt{cie} model, we identify a dominant solution at about $-0.3c$ together with secondary solutions at about $-0.2c$, whose differences could be tentatively interpreted in terms of precession. Future deeper observations will be useful to investigate possible quasi-periodic Doppler shift variations, particularly in light of the quasi-periodic flux modulation reported by \citet{gurpide}.
%However, the higher significance of the time-averaged spectrum and the narrow observed line widths argue against strong variability.

Our findings do not allow us to constrain the nature of the compact object. The analogy with the source SS433 may suggest a black hole accretor.
Alternatively, the two observed CIE plasma components can be explained in a scenario involving a neutron star accretor with a moderately strong magnetic field.
In this context, the accreting material is channelled along the magnetic field lines onto the neutron star’s surface. The combined effect of strong gravitational forces and intense radiation pressure leads to the formation of shocks close to the surface of the NS (see, e.g. \citealt{Inoue2024}), resulting in collisional-ionisation of the plasma. The extreme flux of the super-Eddington NS surface would then result in radiation pressure onto the plasma and, therefore, a (bi)conical outflow similar to a jet (see also \citealt{Abarca2018,Kayanikhoo2025}). 
Radiative magnetohydrodynamic simulations show that, when the thermal pressure within the accretion column exceeds the magnetic pressure in the case of super-Eddington accretion, the material can either be accreted or escape; in the latter case, if channelled along open magnetic field lines, it may be accelerated to relativistic velocities \citep{abolmasov2023}. In this case, the outflows might be partly collimated along the magnetic axes of the NS, as the ejected material has to escape from the magnetosphere without colliding with the magnetospheric flow moving towards the compact object.
Alternatively, radiation-driven mass losses are possible from the magnetospheric flow since the radiation pressure perpendicular to the magnetic field lines can exceed the local magnetic field pressure at high accretion rates and luminosities, and local disruption of the accretion flow is expected \citep{flexer2024}. In the case of a NS accretor in the propeller regime, both a jet ($v\sim0.4-0.5 c$) and a conical wind ($v\sim0.03-0.1 c$) powered by the NS magnetic field can be present simultaneously \citep{romanova2009}. The outflow detected in NGC 5204 X-1 has a line-of-sight velocity of about $0.3c$, placing it midway between a jet and a conical wind. Additionally, mildly-relativistic speeds of the jets may be evidence of systems hosting neutron stars, as found in \citet{russell2024}, because higher speeds are typically measured for black hole systems ($\beta> 0.6$; e.g. \citealt{saikia2019}).

%The \texttt{cie} model scan was also applied to the individual observations. To summarise the results for the different datasets, we present them in Figure \ref{fig:CIE_results}. Specifically, we detected different phases of the plasma in the individual observations. For observation 0921360101, when the source is in the intermediate / hard regime, we detected a rest-frame component with a higher temperature compared to that observed in the archive-only and time-averaged data, along with a fast ($\sim-0.2c$) outflow with a temperature of $\sim1$ keV. For observation 0921360201, when the source is in the unobscured bright / soft regime, we found only a fast ($\sim0.2c$) component with a slightly lower temperature, moving in opposite direction to the one observed in the other observation. These two fast components are likely degenerate with those exhibiting a line-of-sight velocity of $0.3c$.
 
%\begin{figure}
%   \centering
%   \includegraphics[scale=0.5]{Plots/results_CIE_1.pdf}
%   \caption{Best-fit results obtained using the \texttt{cie} model for the different datasets. $L_\mathrm{X}$ is computed in the $0.3-10$ keV energy range.}
%    \label{fig:CIE_results}
%\end{figure}
We also tested photoionised plasma models in both emission and absorption.
In the archive-only data, we do not find detections with a confidence level of at least $3\sigma$. Including the latest observations and analysing the time-averaged spectra, we find evidence of a slow-moving plasma with a line-of-sight velocity of about \textcolor{black}{$4200$ km/s} and low temperature (T $\sim1.5\times10^4$K) at $3 \sigma$, consistent with the results from the analysis of the rest-frame O VII (see Section \ref{sec:ovii}). Its low temperature and velocity are compatible with a thermal wind that is expected to originate in the outer parts of the accretion disc \citep{thermal_winds}.

Using the definition of the ionisation parameter and assuming a photoionised line-emitting plasma, we can obtain a constraint onto the distance from the ionising source, i.e. the inner portion of the accretion disc, $R=\sqrt{L_{ion}/n_\mathrm{H} \xi}$.
Assuming a volume density $n_\mathrm{H}=3\times10^{11} \text{cm}^{-3}$, estimated from the analysis of the O VII triplet, and an ionisation parameter $\xi\sim1.58$ erg/s cm, corresponding to the approximately rest-frame solution obtained from the scan of the time-averaged spectra with the \texttt{pion} model, we estimate a distance $R\sim7$ AU. If we adopt a volume density $n_\mathrm{H}=10^{12} \text{cm}^{-3}$ for the same value of $\xi$, we obtain a distance $R\sim3$ AU. 
Assuming a volume density $n_\mathrm{H}=3\times10^{11} \text{cm}^{-3}$ and an ionisation parameter $\xi\sim500$ erg/s cm, corresponding to the plasma outflow at about $-0.2c$ resulting from the scan of the time-averaged spectra, we estimate a distance $R\sim0.4$ AU. This is consistent with the result obtained for the ULX NGC 1313 X-1 \citep{1313x1_pi} and suggests that the plasma must be located in the very vicinity of the ULX. 

Applying a photoionised absorption model, no detections above $3\sigma$ are found for the archive-only, time-averaged data, or observation 0921360201, while in observation 0921360101 we detect a plasma outflow at $\sim-0.05c$ with low temperature (T $\sim1.9\times10^4$K) at about $3\sigma$. 
%This feature was observed in the archive-only data with lower significance, although it was identified as plasma emitting rather than absorbing, with a lower ionisation. 
This component is consistent with that one found in the broadened-disc state of the ULX NGC 1313 X-1 \citep{1313x1_pi}.

\section{Conclusions} \label{sec:conclusions}
In this work, we analysed XMM-\textit{Newton} data from new and past observations of the ultraluminous X-ray source NGC 5204 X-1, with the aim of studying the geometry of the accretion disc and the outflows. 
From the spectral analysis of spectral stacks and the use of physical model grids of line-emitting and line-absorbing plasma, we detect at high significance a collisionally-ionised plasma blueshifted to $0.3c$ ($>3\sigma$) and another component outflowing in opposite direction, as it is redshifted to $0.3c$ ($>4\sigma$). These detections reveal a biconical structure for the outflow in NGC 5204 X-1, in analogy with the Galactic super-Eddington accretor SS433.
Furthermore, at lower significance, we detect a photoionised plasma in emission moving at about $4000$ km/s towards the observer, likely evidence of a thermal wind. 
The cone of CIE outflowing plasma might suggest super-Eddington accretion (and shocks) onto a black hole or a magnetised neutron star powering this ULX.

%%%%%%%%%%%%%%%%%%%%%%%%%%%%%%%%%%%%%%%%%%%%%%%%%%
\section*{Data Availability}

All data and software used in this work are publicly available from the XMM-Newton Science Archive (https://www.cosmos.esa.int/web/xmm-newton/xsa) and NASA’s HEASARC archive (https://heasarc.gsfc.nasa.gov/). Our codes are publicly available and can be found on GitHub (https://github.com/ciropinto1982).

\begin{acknowledgements}
We acknowledge funding from PRIN MUR 2022 SEAWIND 2022Y2T94C, supported by European Union - Next Generation EU, Mission 4 Component 1 CUP C53D23001330006, INAF Large Grant 2023 BLOSSOM O.F. 1.05.23.01.13 and INAF Large Grant 2024 "Timing the Ultra Luminous X-ray Pulsars (TULiP)". This work has been partially supported by the ASI-INAF program I/004/11/6. DJW acknowledges support from the STFC grant code ST/Y001060/1.
The authors thank Alexander Mushtukov for useful discussion on outflows from the NS and the magnetosphere.
\end{acknowledgements}

\bibliographystyle{aa}
\bibliography{ref.bib}

@ARTICLE{Inoue2024,
       author = {{Inoue}, Akihiro and {Ohsuga}, Ken and {Takahashi}, Hiroyuki R. and {Asahina}, Yuta and {Middleton}, Matthew J.},
        title = "{GR-RMHD Simulations of Super-Eddington Accretion Flows onto a Neutron Star with Dipole and Quadrupole Magnetic Fields}",
      journal = {\apj},
         year = 2024,
        month = dec,
       volume = {977},
       number = {1},
          eid = {10},
        pages = {10},
          doi = {10.3847/1538-4357/ad8885},
archivePrefix = {arXiv},
       eprint = {2410.17707},
 primaryClass = {astro-ph.HE},
       adsurl = {https://ui.adsabs.harvard.edu/abs/2024ApJ...977...10I}
}

@ARTICLE{Abarca2018,
       author = {{Abarca}, David and {Klu{\'z}niak}, W{\l}odek and {S{\k{a}}dowski}, Aleksander},
        title = "{Radiative GRMHD simulations of accretion and outflow in non-magnetized neutron stars and ultraluminous X-ray sources}",
      journal = {\mnras},
         year = 2018,
        month = sep,
       volume = {479},
       number = {3},
        pages = {3936-3951},
          doi = {10.1093/mnras/sty1602},
archivePrefix = {arXiv},
       eprint = {1712.08295},
 primaryClass = {astro-ph.HE},
       adsurl = {https://ui.adsabs.harvard.edu/abs/2018MNRAS.479.3936A}
}

@ARTICLE{Kayanikhoo2025,
       author = {{Kayanikhoo}, Fatemeh and {Klu{\'z}niak}, W{\l}odek and {{\v{C}}emelji{\'c}}, Miljenko},
        title = "{ULX Collimation by Outflows in Moderately Magnetized Neutron Stars}",
      journal = {\apj},
         year = 2025,
        month = apr,
       volume = {982},
       number = {2},
          eid = {95},
        pages = {95},
          doi = {10.3847/1538-4357/adb714},
archivePrefix = {arXiv},
       eprint = {2502.12282},
 primaryClass = {astro-ph.HE},
       adsurl = {https://ui.adsabs.harvard.edu/abs/2025ApJ...982...95K}
}

@ARTICLE{King2023,
       author = {{King}, Andrew and {Lasota}, Jean-Pierre and {Middleton}, Matthew},
        title = "{Ultraluminous X-ray sources}",
      journal = {\nar},
         year = 2023,
        month = jun,
       volume = {96},
          eid = {101672},
        pages = {101672},
          doi = {10.1016/j.newar.2022.101672},
archivePrefix = {arXiv},
       eprint = {2302.10605},
 primaryClass = {astro-ph.HE},
       adsurl = {https://ui.adsabs.harvard.edu/abs/2023NewAR..9601672K}
}

@ARTICLE{Kosec2021,
       author = {{Kosec}, P. and {Pinto}, C. and {Reynolds}, C.~S. and {Guainazzi}, M. and {Kara}, E. and {Walton}, D.~J. and {Fabian}, A.~C. and {Parker}, M.~L. and {Valtchanov}, I.},
        title = "{Ionized emission and absorption in a large sample of ultraluminous X-ray sources}",
      journal = {\mnras},
         year = 2021,
        month = dec,
       volume = {508},
       number = {3},
        pages = {3569-3588},
           doi = {10.1093/mnras/stab2856},
archivePrefix = {arXiv},
       eprint = {2109.14683},
 primaryClass = {astro-ph.HE},
       adsurl = {https://ui.adsabs.harvard.edu/abs/2021MNRAS.508.3569K}
}

@ARTICLE{Pinto2020b,
       author = {{Pinto}, C. and others},
        title = "{XMM-Newton campaign on ultraluminous X-ray source NGC 1313 X-1: wind versus state variability}",
      journal = {\mnras},
         year = 2020,
        month = mar,
       volume = {492},
       number = {4},
        pages = {4646-4665},
            doi = {10.1093/mnras/staa118},
archivePrefix = {arXiv},
       eprint = {1911.09568},
 primaryClass = {astro-ph.HE},
       adsurl = {https://ui.adsabs.harvard.edu/abs/2020MNRAS.492.4646P}
}

@Inbook{Pinto2023a,
author="Pinto, Ciro
and Walton, Dominic J.",
editor="Bambi, Cosimo
and Jiang, Jiachen",
title="Ultra-Luminous X-Ray Sources: Extreme Accretion and Feedback",
bookTitle="High-Resolution X-ray Spectroscopy: Instrumentation, Data Analysis, and Science",
year="2023",
publisher="Springer Nature Singapore",
address="Singapore",
pages="345--391",
isbn="978-981-99-4409-5",
doi="10.1007/978-981-99-4409-5_12",
url="https://doi.org/10.1007/978-981-99-4409-5_12"
}

@ARTICLE{Poutanen2007,
   author = {{Poutanen}, J. and {Lipunova}, G. and {Fabrika}, S. and {Butkevich}, A.~G. and 
	{Abolmasov}, P.},
    title = "{Supercritically accreting stellar mass black holes as ultraluminous X-ray sources}",
  journal = {\mnras},
   eprint = {astro-ph/0609274},
     year = 2007,
    month = may,
   volume = 377,
    pages = {1187-1194},
        doi = {10.1111/j.1365-2966.2007.11668.x},
   adsurl = {http://adsabs.harvard.edu/abs/2007MNRAS.377.1187P}
}

@ARTICLE{pulx_swift,
       author = {{van den Eijnden}, J. and {Degenaar}, N. and {Schulz}, N.~S. and {Nowak}, M.~A. and {Wijnands}, R. and {Russell}, T.~D. and {Hern{\'a}ndez Santisteban}, J.~V. and {Bahramian}, A. and {Maccarone}, T.~J. and {Kennea}, J.~A. and {Heinke}, C.~O.},
        title = "{Chandra reveals a possible ultrafast outflow in the super-Eddington Be/X-ray binary Swift J0243.6+6124}",
      journal = {\mnras},
         year = 2019,
        month = aug,
       volume = {487},
       number = {3},
        pages = {4355-4371},
          doi = {10.1093/mnras/stz1548},
archivePrefix = {arXiv},
       eprint = {1906.01597},
 primaryClass = {astro-ph.HE},
       adsurl = {https://ui.adsabs.harvard.edu/abs/2019MNRAS.487.4355V}
}

@ARTICLE{ss433_lines,
       author = {{Marshall}, Herman L. and {Canizares}, Claude R. and {Schulz}, Norbert S.},
        title = "{The High-Resolution X-Ray Spectrum of SS 433 Using the Chandra HETGS}",
      journal = {\apj},
         year = 2002,
        month = jan,
       volume = {564},
       number = {2},
        pages = {941-952},
          doi = {10.1086/324398},
archivePrefix = {arXiv},
       eprint = {astro-ph/0108206},
 primaryClass = {astro-ph},
       adsurl = {https://ui.adsabs.harvard.edu/abs/2002ApJ...564..941M}
}

@ARTICLE{Bachetti2013,
   author = {{Bachetti}, M. and {Rana}, V. and {Walton}, D.~J. and {Barret}, D. and 
	{Harrison}, F.~A. and others},
    title = "{The Ultraluminous X-Ray Sources NGC 1313 X-1 and X-2: A Broadband Study with NuSTAR and XMM-Newton}",
  journal = {\apj},
archivePrefix = "arXiv",
   eprint = {1310.0745},
 primaryClass = "astro-ph.HE",
     year = 2013,
    month = dec,
   volume = 778,
      eid = {163},
    pages = {163},
        doi = {10.1088/0004-637X/778/2/163},
   adsurl = {http://adsabs.harvard.edu/abs/2013ApJ...778..163B}
}

@ARTICLE{m82x2,
       author = {{Bachetti}, M. and {Harrison}, F.~A. and {Walton}, D.~J. and {Grefenstette}, B.~W. and {Chakrabarty}, D. and {F{\"u}rst}, F. and {Barret}, D. and {Beloborodov}, A. and {Boggs}, S.~E. and {Christensen}, F.~E. and {Craig}, W.~W. and {Fabian}, A.~C. and {Hailey}, C.~J. and {Hornschemeier}, A. and {Kaspi}, V. and {Kulkarni}, S.~R. and {Maccarone}, T. and {Miller}, J.~M. and {Rana}, V. and {Stern}, D. and {Tendulkar}, S.~P. and {Tomsick}, J. and {Webb}, N.~A. and {Zhang}, W.~W.},
        title = "{An ultraluminous X-ray source powered by an accreting neutron star}",
      journal = {\nat},
         year = 2014,
        month = oct,
       volume = {514},
       number = {7521},
        pages = {202-204},
          doi = {10.1038/nature13791},
archivePrefix = {arXiv},
       eprint = {1410.3590},
 primaryClass = {astro-ph.HE},
       adsurl = {https://ui.adsabs.harvard.edu/abs/2014Natur.514..202B}
}

@ARTICLE{ulxstate,
       author = {{Gladstone}, Jeanette C. and {Roberts}, Timothy P. and {Done}, Chris},
        title = "{The ultraluminous state}",
      journal = {\mnras},
         year = 2009,
        month = aug,
       volume = {397},
       number = {4},
        pages = {1836-1851},
          doi = {10.1111/j.1365-2966.2009.15123.x},
archivePrefix = {arXiv},
       eprint = {0905.4076},
 primaryClass = {astro-ph.CO},
       adsurl = {https://ui.adsabs.harvard.edu/abs/2009MNRAS.397.1836G}
}

@INPROCEEDINGS{spex,
       author = {{Kaastra}, J.~S. and {Mewe}, R. and {Nieuwenhuijzen}, H.},
        title = "{SPEX: a new code for spectral analysis of X \& UV spectra.}",
    booktitle = {UV and X-ray Spectroscopy of Astrophysical and Laboratory Plasmas},
         year = 1996,
       editor = {{Yamashita}, K. and {Watanabe}, T.},
        month = jan,
        pages = {411-414},
       adsurl = {https://ui.adsabs.harvard.edu/abs/1996uxsa.conf..411K}
}

@ARTICLE{cash,
       author = {{Cash}, W.},
        title = "{Parameter estimation in astronomy through application of the likelihood ratio.}",
      journal = {\apj},
         year = 1979,
        month = mar,
       volume = {228},
        pages = {939-947},
          doi = {10.1086/156922},
       adsurl = {https://ui.adsabs.harvard.edu/abs/1979ApJ...228..939C}
}

@ARTICLE{gurpide,
       author = {{G{\'u}rpide}, A. and {Godet}, O. and {Vasilopoulos}, G. and {Webb}, N.~A. and {Olive}, J. -F.},
        title = "{Discovery of a recurrent spectral evolutionary cycle in the ultra-luminous X-ray sources Holmberg II X-1 and NGC 5204 X-1}",
      journal = {\aap},
         year = 2021,
        month = oct,
       volume = {654},
          eid = {A10},
        pages = {A10},
          doi = {10.1051/0004-6361/202140781},
archivePrefix = {arXiv},
       eprint = {2106.05708},
 primaryClass = {astro-ph.HE},
       adsurl = {https://ui.adsabs.harvard.edu/abs/2021A&A...654A..10G}
}

@ARTICLE{Krolik1981,
   author = {{Krolik}, J.~H. and {McKee}, C.~F. and {Tarter}, C.~B.},
    title = "{Two-phase models of quasar emission line regions}",
  journal = {\apj},
     year = 1981,
    month = oct,
   volume = 249,
    pages = {422-442},
        doi = {10.1086/159303},
   adsurl = {http://adsabs.harvard.edu/abs/1981ApJ...249..422K}
}

@ARTICLE{gr_ratios,
       author = {{Porquet}, D. and {Dubau}, J.},
        title = "{X-ray photoionized plasma diagnostics with helium-like ions. Application to warm absorber-emitter in active galactic nuclei}",
      journal = {\aaps},
         year = 2000,
        month = may,
       volume = {143},
        pages = {495-514},
          doi = {10.1051/aas:2000192},
archivePrefix = {arXiv},
       eprint = {astro-ph/0002319},
 primaryClass = {astro-ph},
       adsurl = {https://ui.adsabs.harvard.edu/abs/2000A&AS..143..495P}
}

@ARTICLE{luminosity_5204,
       author = {{Mukherjee}, E.~S. and {Walton}, D.~J. and {Bachetti}, M. and {Harrison}, F.~A. and {Barret}, D. and {Bellm}, E. and {Boggs}, S.~E. and {Christensen}, F.~E. and {Craig}, W.~W. and {Fabian}, A.~C. and {Fuerst}, F. and {Grefenstette}, B.~W. and {Hailey}, C.~J. and {Madsen}, K.~K. and {Middleton}, M.~J. and {Miller}, J.~M. and {Rana}, V. and {Stern}, D. and {Zhang}, W.},
        title = "{A Hard X-Ray Study of the Ultraluminous X-Ray Source NGC 5204 X-1 with NuSTAR and XMM-Newton}",
      journal = {\apj},
         year = 2015,
        month = jul,
       volume = {808},
       number = {1},
          eid = {64},
        pages = {64},
          doi = {10.1088/0004-637X/808/1/64},
archivePrefix = {arXiv},
       eprint = {1502.01764},
 primaryClass = {astro-ph.HE},
       adsurl = {https://ui.adsabs.harvard.edu/abs/2015ApJ...808...64M}
}

@ARTICLE{Mehdipour2016,
   author = {{Mehdipour}, M. and {Kaastra}, J.~S. and {Kallman}, T.},
    title = "{Systematic comparison of photoionised plasma codes with application to spectroscopic studies of AGN in X-rays}",
  journal = {\aap},
archivePrefix = "arXiv",
   eprint = {1610.03080},
 primaryClass = "astro-ph.HE",
     year = 2016,
    month = dec,
   volume = 596,
      eid = {A65},
    pages = {A65},
        doi = {10.1051/0004-6361/201628721},
   adsurl = {http://adsabs.harvard.edu/abs/2016A%26A...596A..65M}
}

@ARTICLE{Middleton2015a,
   author = {{Middleton}, M.~J. and {Heil}, L. and {Pintore}, F. and {Walton}, D.~J. and 
	{Roberts}, T.~P.},
    title = "{A spectral-timing model for ULXs in the supercritical regime}",
  journal = {\mnras},
archivePrefix = "arXiv",
   eprint = {1412.4532},
 primaryClass = "astro-ph.HE",
     year = 2015,
    month = mar,
   volume = 447,
    pages = {3243-3263},
        doi = {10.1093/mnras/stu2644},
   adsurl = {http://adsabs.harvard.edu/abs/2015MNRAS.447.3243M}
}

@ARTICLE{supercritical_accr,
       author = {{Ohsuga}, Ken and {Mori}, Masao and {Nakamoto}, Taishi and {Mineshige}, Shin},
        title = "{Supercritical Accretion Flows around Black Holes: Two-dimensional, Radiation Pressure-dominated Disks with Photon Trapping}",
      journal = {\apj},
         year = 2005,
        month = jul,
       volume = {628},
       number = {1},
        pages = {368-381},
          doi = {10.1086/430728},
archivePrefix = {arXiv},
       eprint = {astro-ph/0504168},
 primaryClass = {astro-ph},
       adsurl = {https://ui.adsabs.harvard.edu/abs/2005ApJ...628..368O}
}

@ARTICLE{uls,
       author = {{Sutton}, Andrew D. and {Roberts}, Timothy P. and {Middleton}, Matthew J.},
        title = "{The ultraluminous state revisited: fractional variability and spectral shape as diagnostics of super-Eddington accretion}",
      journal = {\mnras},
         year = 2013,
        month = oct,
       volume = {435},
       number = {2},
        pages = {1758-1775},
          doi = {10.1093/mnras/stt1419},
archivePrefix = {arXiv},
       eprint = {1307.8044},
 primaryClass = {astro-ph.HE},
       adsurl = {https://ui.adsabs.harvard.edu/abs/2013MNRAS.435.1758S}
}

@ARTICLE{1313x1_mid,
       author = {{Middleton}, Matthew J. and {Walton}, Dominic J. and {Fabian}, Andrew and {Roberts}, Timothy P. and {Heil}, Lucy and {Pinto}, Ciro and {Anderson}, Gemma and {Sutton}, Andrew},
        title = "{Diagnosing the accretion flow in ultraluminous X-ray sources using soft X-ray atomic features}",
      journal = {\mnras},
         year = 2015,
        month = dec,
       volume = {454},
       number = {3},
        pages = {3134-3142},
          doi = {10.1093/mnras/stv2214},
archivePrefix = {arXiv},
       eprint = {1509.06760},
 primaryClass = {astro-ph.HE},
       adsurl = {https://ui.adsabs.harvard.edu/abs/2015MNRAS.454.3134M}
}

@ARTICLE{1313x1_pi,
       author = {{Pinto}, C. and {Walton}, D.~J. and {Kara}, E. and {Parker}, M.~L. and {Soria}, R. and {Kosec}, P. and {Middleton}, M.~J. and {Alston}, W.~N. and {Fabian}, A.~C. and {Guainazzi}, M. and {Roberts}, T.~P. and {Fuerst}, F. and {Earnshaw}, H.~P. and {Sathyaprakash}, R. and {Barret}, D.},
        title = "{XMM-Newton campaign on ultraluminous X-ray source NGC 1313 X-1: wind versus state variability}",
      journal = {\mnras},
         year = 2020,
        month = mar,
       volume = {492},
       number = {4},
        pages = {4646-4665},
          doi = {10.1093/mnras/staa118},
archivePrefix = {arXiv},
       eprint = {1911.09568},
 primaryClass = {astro-ph.HE},
       adsurl = {https://ui.adsabs.harvard.edu/abs/2020MNRAS.492.4646P}
}

@ARTICLE{thermal_winds,
       author = {{Middleton}, Matthew J. and {Higginbottom}, Nick and {Knigge}, Christian and {Khan}, Norman and {Wiktorowicz}, Grzegorz},
        title = "{Thermally driven winds in ultraluminous X-ray sources}",
      journal = {\mnras},
         year = 2022,
        month = jan,
       volume = {509},
       number = {1},
        pages = {1119-1126},
          doi = {10.1093/mnras/stab2991},
archivePrefix = {arXiv},
       eprint = {2110.08249},
 primaryClass = {astro-ph.HE},
       adsurl = {https://ui.adsabs.harvard.edu/abs/2022MNRAS.509.1119M}
}

@ARTICLE{spectral_lines1,
       author = {{Pinto}, Ciro and {Middleton}, Matthew J. and {Fabian}, Andrew C.},
        title = "{Resolved atomic lines reveal outflows in two ultraluminous X-ray sources}",
      journal = {\nat},
         year = 2016,
        month = may,
       volume = {533},
       number = {7601},
        pages = {64-67},
          doi = {10.1038/nature17417},
archivePrefix = {arXiv},
       eprint = {1604.08593},
 primaryClass = {astro-ph.HE},
       adsurl = {https://ui.adsabs.harvard.edu/abs/2016Natur.533...64P}
}

@ARTICLE{spectral_lines2,
       author = {{Pinto}, C. and {Alston}, W. and {Soria}, R. and {Middleton}, M.~J. and {Walton}, D.~J. and {Sutton}, A.~D. and {Fabian}, A.~C. and {Earnshaw}, H. and {Urquhart}, R. and {Kara}, E. and {Roberts}, T.~P.},
        title = "{From ultraluminous X-ray sources to ultraluminous supersoft sources: NGC 55 ULX, the missing link}",
      journal = {\mnras},
         year = 2017,
        month = jul,
       volume = {468},
       number = {3},
        pages = {2865-2883},
          doi = {10.1093/mnras/stx641},
archivePrefix = {arXiv},
       eprint = {1612.05569},
 primaryClass = {astro-ph.HE},
       adsurl = {https://ui.adsabs.harvard.edu/abs/2017MNRAS.468.2865P}
}

@ARTICLE{imbhs,
       author = {{Colbert}, Edward J.~M. and {Mushotzky}, Richard F.},
        title = "{The Nature of Accreting Black Holes in Nearby Galaxy Nuclei}",
      journal = {\apj},
         year = 1999,
        month = jul,
       volume = {519},
       number = {1},
        pages = {89-107},
          doi = {10.1086/307356},
archivePrefix = {arXiv},
       eprint = {astro-ph/9901023},
 primaryClass = {astro-ph},
       adsurl = {https://ui.adsabs.harvard.edu/abs/1999ApJ...519...89C}
}

@ARTICLE{Walton2018,
       author = {{Walton}, D.~J. and {F{\"u}rst}, F. and {Heida}, M. and {Harrison}, F.~A. and {Barret}, D. and {Stern}, D. and {Bachetti}, M. and {Brightman}, M. and {Fabian}, A.~C. and {Middleton}, M.~J.},
        title = "{Evidence for Pulsar-like Emission Components in the Broadband ULX Sample}",
      journal = {\apj},
         year = 2018,
        month = apr,
       volume = {856},
       number = {2},
          eid = {128},
        pages = {128},
          doi = {10.3847/1538-4357/aab610},
archivePrefix = {arXiv},
       eprint = {1803.04424},
 primaryClass = {astro-ph.HE},
       adsurl = {https://ui.adsabs.harvard.edu/abs/2018ApJ...856..128W}
}

@ARTICLE{miller2015,
       author = {{Miller}, Jon M. and {Kaastra}, Jelle S. and {Miller}, M. Coleman and {Reynolds}, Mark T. and {Brown}, Gregory and {Cenko}, S. Bradley and {Drake}, Jeremy J. and {Gezari}, Suvi and {Guillochon}, James and {Gultekin}, Kayhan and {Irwin}, Jimmy and {Levan}, Andrew and {Maitra}, Dipankar and {Maksym}, W. Peter and {Mushotzky}, Richard and {O'Brien}, Paul and {Paerels}, Frits and {de Plaa}, Jelle and {Ramirez-Ruiz}, Enrico and {Strohmayer}, Tod and {Tanvir}, Nial},
        title = "{Flows of X-ray gas reveal the disruption of a star by a massive black hole}",
      journal = {\nat},
         year = 2015,
        month = oct,
       volume = {526},
       number = {7574},
        pages = {542-545},
          doi = {10.1038/nature15708},
archivePrefix = {arXiv},
       eprint = {1510.06348},
 primaryClass = {astro-ph.HE},
       adsurl = {https://ui.adsabs.harvard.edu/abs/2015Natur.526..542M}
}

@ARTICLE{abolmasov2023,
       author = {{Abolmasov}, Pavel and {Lipunova}, Galina},
        title = "{Simulating the shock dynamics of a neutron star accretion column}",
      journal = {\mnras},
         year = 2023,
        month = sep,
       volume = {524},
       number = {3},
        pages = {4148-4167},
          doi = {10.1093/mnras/stad1951},
archivePrefix = {arXiv},
       eprint = {2207.12312},
 primaryClass = {astro-ph.HE},
       adsurl = {https://ui.adsabs.harvard.edu/abs/2023MNRAS.524.4148A}
}

@ARTICLE{flexer2024,
       author = {{Flexer}, Caitlyn and {Mushtukov}, Alexander A.},
        title = "{Coupling of radiation and magnetospheric accretion flow in ULX pulsars: radiation pressure and photon escape time}",
      journal = {\mnras},
         year = 2024,
        month = apr,
       volume = {529},
       number = {2},
        pages = {1571-1578},
          doi = {10.1093/mnras/stae653},
archivePrefix = {arXiv},
       eprint = {2403.07713},
 primaryClass = {astro-ph.HE},
       adsurl = {https://ui.adsabs.harvard.edu/abs/2024MNRAS.529.1571F}
}

@ARTICLE{marshall2013,
       author = {{Marshall}, Herman L. and {Canizares}, Claude R. and {Hillwig}, Todd and {Mioduszewski}, Amy and {Rupen}, Michael and {Schulz}, Norbert S. and {Nowak}, Michael and {Heinz}, Sebastian},
        title = "{Multiwavelength Observations of the SS 433 Jets}",
      journal = {\apj},
         year = 2013,
        month = sep,
       volume = {775},
       number = {1},
          eid = {75},
        pages = {75},
          doi = {10.1088/0004-637X/775/1/75},
archivePrefix = {arXiv},
       eprint = {1307.8427},
 primaryClass = {astro-ph.HE},
       adsurl = {https://ui.adsabs.harvard.edu/abs/2013ApJ...775...75M}
}

@ARTICLE{lodders2009,
       author = {{Lodders}, K. and {Palme}, H. and {Gail}, H. -P.},
        title = "{Abundances of the Elements in the Solar System}",
      journal = {Landolt B{\"o}rnstein},
         year = 2009,
        month = jan,
       volume = {4B},
        pages = {712},
          doi = {10.1007/978-3-540-88055-4_34},
archivePrefix = {arXiv},
       eprint = {0901.1149},
 primaryClass = {astro-ph.EP},
       adsurl = {https://ui.adsabs.harvard.edu/abs/2009LanB...4B..712L}
}

@ARTICLE{takeuchi2013,
       author = {{Takeuchi}, Shun and {Ohsuga}, Ken and {Mineshige}, Shin},
        title = "{Clumpy Outflows from Supercritical Accretion Flow}",
      journal = {\pasj},
         year = 2013,
        month = aug,
       volume = {65},
          eid = {88},
        pages = {88},
          doi = {10.1093/pasj/65.4.88},
archivePrefix = {arXiv},
       eprint = {1305.1023},
 primaryClass = {astro-ph.HE},
       adsurl = {https://ui.adsabs.harvard.edu/abs/2013PASJ...65...88T}
}

@software{kaastra2022,
       author = {{Kaastra}, J.~S. and {Raassen}, A.~J.~J. and {de Plaa}, J. and {Gu}, Liyi},
        title = "{SPEX X-ray spectral fitting package}",
         year = 2022,
        month = aug,
          eid = {10.5281/zenodo.7037609},
          doi = {10.5281/zenodo.7037609},
      version = {3.07.01},
    publisher = {Zenodo},
       adsurl = {https://ui.adsabs.harvard.edu/abs/2022zndo...7037609K}
}

@ARTICLE{evans2009,
       author = {{Evans}, P.~A. and {Beardmore}, A.~P. and {Page}, K.~L. and {Osborne}, J.~P. and {O'Brien}, P.~T. and {Willingale}, R. and {Starling}, R.~L.~C. and {Burrows}, D.~N. and {Godet}, O. and {Vetere}, L. and {Racusin}, J. and {Goad}, M.~R. and {Wiersema}, K. and {Angelini}, L. and {Capalbi}, M. and {Chincarini}, G. and {Gehrels}, N. and {Kennea}, J.~A. and {Margutti}, R. and {Morris}, D.~C. and {Mountford}, C.~J. and {Pagani}, C. and {Perri}, M. and {Romano}, P. and {Tanvir}, N.},
        title = "{Methods and results of an automatic analysis of a complete sample of Swift-XRT observations of GRBs}",
      journal = {\mnras},
         year = 2009,
        month = aug,
       volume = {397},
       number = {3},
        pages = {1177-1201},
          doi = {10.1111/j.1365-2966.2009.14913.x},
archivePrefix = {arXiv},
       eprint = {0812.3662},
 primaryClass = {astro-ph},
       adsurl = {https://ui.adsabs.harvard.edu/abs/2009MNRAS.397.1177E}
}

@ARTICLE{dickey1990,
       author = {{Dickey}, John M. and {Lockman}, Felix J.},
        title = "{H I in the galaxy.}",
      journal = {\araa},
         year = 1990,
        month = jan,
       volume = {28},
        pages = {215-261},
          doi = {10.1146/annurev.aa.28.090190.001243},
       adsurl = {https://ui.adsabs.harvard.edu/abs/1990ARA&A..28..215D}
}

@ARTICLE{Kosec2018b,
       author = {{Kosec}, P. and {Pinto}, C. and {Walton}, D.~J. and {Fabian}, A.~C. and {Bachetti}, M. and {Brightman}, M. and {F{\"u}rst}, F. and {Grefenstette}, B.~W.},
        title = "{Evidence for a variable Ultrafast Outflow in the newly discovered Ultraluminous Pulsar NGC 300 ULX-1}",
      journal = {\mnras},
         year = 2018,
        month = sep,
       volume = {479},
       number = {3},
        pages = {3978-3986},
          doi = {10.1093/mnras/sty1626},
archivePrefix = {arXiv},
       eprint = {1803.02367},
 primaryClass = {astro-ph.HE},
       adsurl = {https://ui.adsabs.harvard.edu/abs/2018MNRAS.479.3978K}
}

@ARTICLE{Kosec2018a,
   author = {{Kosec}, P. and {Pinto}, C. and {Fabian}, A.~C. and {Walton}, D.~J.
	},
    title = "{Searching for outflows in ultraluminous X-ray sources through high-resolution X-ray spectroscopy}",
  journal = {\mnras},
archivePrefix = "arXiv",
   eprint = {1710.06438},
 primaryClass = "astro-ph.HE",
     year = 2018,
    month = feb,
   volume = 473,
    pages = {5680-5697},
        doi = {10.1093/mnras/stx2695},
   adsurl = {http://adsabs.harvard.edu/abs/2018MNRAS.473.5680K}
}

@ARTICLE{Pinto2021,
       author = {{Pinto}, C. and {Soria}, R. and {Walton}, D.~J. and {D'A{\`\i}}, A. and {Pintore}, F. and {Kosec}, P. and {Alston}, W.~N. and {Fuerst}, F. and {Middleton}, M.~J. and {Roberts}, T.~P. and et al.},
        title = "{XMM-Newton campaign on the ultraluminous X-ray source NGC 247 ULX-1: outflows}",
      journal = {\mnras},
         year = 2021,
        month = aug,
       volume = {505},
       number = {4},
        pages = {5058-5074},
          doi = {10.1093/mnras/stab1648},
archivePrefix = {arXiv},
       eprint = {2104.11164},
 primaryClass = {astro-ph.HE},
       adsurl = {https://ui.adsabs.harvard.edu/abs/2021MNRAS.505.5058P}
}

@ARTICLE{Walton2016,
       author = {{Walton}, D.~J. and {Middleton}, M.~J. and {Pinto}, C. and {Fabian}, A.~C. and {Bachetti}, M. and {Barret}, D. and {Brightman}, M. and {Fuerst}, F. and {Harrison}, F.~A. and {Miller}, J.~M. and et al.},
        title = "{An Iron K Component to the Ultrafast Outflow in NGC 1313 X-1}",
      journal = {\apjl},
         year = 2016,
        month = aug,
       volume = {826},
       number = {2},
          eid = {L26},
        pages = {L26},
          doi = {10.3847/2041-8205/826/2/L26},
archivePrefix = {arXiv},
       eprint = {1607.03124},
 primaryClass = {astro-ph.HE},
       adsurl = {https://ui.adsabs.harvard.edu/abs/2016ApJ...826L..26W}
}

@ARTICLE{distance,
       author = {{Bottinelli}, L. and {Gouguenheim}, L. and {Paturel}, G. and {de Vaucouleurs}, G.},
        title = "{HI line studies of galaxies. III. Distance moduli of 822 disk galaxies.}",
      journal = {\aaps},
         year = 1984,
        month = jun,
       volume = {56},
        pages = {381-413},
       adsurl = {https://ui.adsabs.harvard.edu/abs/1984A&AS...56..381B}
}

@ARTICLE{kaastra2016,
       author = {{Kaastra}, J.~S. and {Bleeker}, J.~A.~M.},
        title = "{Optimal binning of X-ray spectra and response matrix design}",
      journal = {\aap},
         year = 2016,
        month = mar,
       volume = {587},
          eid = {A151},
        pages = {A151},
          doi = {10.1051/0004-6361/201527395},
archivePrefix = {arXiv},
       eprint = {1601.05309},
 primaryClass = {astro-ph.IM},
       adsurl = {https://ui.adsabs.harvard.edu/abs/2016A&A...587A.151K}
}

@ARTICLE{russell2024,
       author = {{Russell}, Thomas D. and {Degenaar}, Nathalie and {van den Eijnden}, Jakob and {Maccarone}, Thomas and {Tetarenko}, Alexandra J. and {S{\'a}nchez-Fern{\'a}ndez}, Celia and {Miller-Jones}, James C.~A. and {Kuulkers}, Erik and {Del Santo}, Melania},
        title = "{Thermonuclear explosions on neutron stars reveal the speed of their jets}",
      journal = {\nat},
         year = 2024,
        month = mar,
       volume = {627},
       number = {8005},
        pages = {763-766},
          doi = {10.1038/s41586-024-07133-5},
archivePrefix = {arXiv},
       eprint = {2403.18135},
 primaryClass = {astro-ph.HE},
       adsurl = {https://ui.adsabs.harvard.edu/abs/2024Natur.627..763R}
}

@ARTICLE{saikia2019,
       author = {{Saikia}, Payaswini and {Russell}, David M. and {Bramich}, D.~M. and {Miller-Jones}, James C.~A. and {Baglio}, Maria Cristina and {Degenaar}, Nathalie},
        title = "{Lorentz Factors of Compact Jets in Black Hole X-Ray Binaries}",
      journal = {\apj},
         year = 2019,
        month = dec,
       volume = {887},
       number = {1},
          eid = {21},
        pages = {21},
          doi = {10.3847/1538-4357/ab4a09},
archivePrefix = {arXiv},
       eprint = {1910.01151},
 primaryClass = {astro-ph.HE},
       adsurl = {https://ui.adsabs.harvard.edu/abs/2019ApJ...887...21S}
}

@ARTICLE{brightman2016,
       author = {{Brightman}, Murray and {Harrison}, Fiona and {Walton}, Dominic J. and {Fuerst}, Felix and {Hornschemeier}, Ann and {Zezas}, Andreas and {Bachetti}, Matteo and {Grefenstette}, Brian and {Ptak}, Andrew and {Tendulkar}, Shriharsh and {Yukita}, Mihoko},
        title = "{Spectral and Temporal Properties of the Ultraluminous X-Ray Pulsar in M82 from 15 years of Chandra Observations and Analysis of the Pulsed Emission Using NuSTAR}",
      journal = {\apj},
         year = 2016,
        month = jan,
       volume = {816},
       number = {2},
          eid = {60},
        pages = {60},
          doi = {10.3847/0004-637X/816/2/60},
archivePrefix = {arXiv},
       eprint = {1507.06014},
 primaryClass = {astro-ph.HE},
       adsurl = {https://ui.adsabs.harvard.edu/abs/2016ApJ...816...60B}
}

@ARTICLE{protassov2002,
       author = {{Protassov}, Rostislav and {van Dyk}, David A. and {Connors}, Alanna and {Kashyap}, Vinay L. and {Siemiginowska}, Aneta},
        title = "{Statistics, Handle with Care: Detecting Multiple Model Components with the Likelihood Ratio Test}",
      journal = {\apj},
         year = 2002,
        month = may,
       volume = {571},
       number = {1},
        pages = {545-559},
          doi = {10.1086/339856},
archivePrefix = {arXiv},
       eprint = {astro-ph/0201547},
 primaryClass = {astro-ph},
       adsurl = {https://ui.adsabs.harvard.edu/abs/2002ApJ...571..545P}
}

@ARTICLE{gu2022,
       author = {{Gu}, Liyi and {Mao}, Junjie and {Kaastra}, Jelle S. and {Mehdipour}, Missagh and {Pinto}, Ciro and {Grafton-Waters}, Sam and {Bianchi}, Stefano and {Landt}, Hermine and {Branduardi-Raymont}, Graziella and {Costantini}, Elisa and {Ebrero}, Jacobo and {Petrucci}, Pierre-Olivier and {Behar}, Ehud and {di Gesu}, Laura and {De Marco}, Barbara and {Matt}, Giorgio and {Mitchell}, Jake A.~J. and {Peretz}, Uria and {Ursini}, Francesco and {Ward}, Martin},
        title = "{Detection of an unidentified soft X-ray emission feature in NGC 5548}",
      journal = {\aap},
         year = 2022,
        month = sep,
       volume = {665},
          eid = {A93},
        pages = {A93},
          doi = {10.1051/0004-6361/202244075},
archivePrefix = {arXiv},
       eprint = {2207.09114},
 primaryClass = {astro-ph.HE},
       adsurl = {https://ui.adsabs.harvard.edu/abs/2022A&A...665A..93G}
}

@ARTICLE{romanova2009,
       author = {{Romanova}, M.~M. and {Ustyugova}, G.~V. and {Koldoba}, A.~V. and {Lovelace}, R.~V.~E.},
        title = "{Launching of conical winds and axial jets from the disc-magnetosphere boundary: axisymmetric and 3D simulations}",
      journal = {\mnras},
         year = 2009,
        month = nov,
       volume = {399},
       number = {4},
        pages = {1802-1828},
          doi = {10.1111/j.1365-2966.2009.15413.x},
archivePrefix = {arXiv},
       eprint = {0907.3394},
 primaryClass = {astro-ph.SR},
       adsurl = {https://ui.adsabs.harvard.edu/abs/2009MNRAS.399.1802R}
}

@ARTICLE{swift,
       author = {{Gehrels}, N. and {Chincarini}, G. and {Giommi}, P. and {Mason}, K.~O. and {Nousek}, J.~A. and {Wells}, A.~A. and {White}, N.~E. and {Barthelmy}, S.~D. and {Burrows}, D.~N. and {Cominsky}, L.~R. and {Hurley}, K.~C. and {Marshall}, F.~E. and {M{\'e}sz{\'a}ros}, P. and {Roming}, P.~W.~A. and {Angelini}, L. and {Barbier}, L.~M. and {Belloni}, T. and {Campana}, S. and {Caraveo}, P.~A. and {Chester}, M.~M. and {Citterio}, O. and {Cline}, T.~L. and {Cropper}, M.~S. and {Cummings}, J.~R. and {Dean}, A.~J. and {Feigelson}, E.~D. and {Fenimore}, E.~E. and {Frail}, D.~A. and {Fruchter}, A.~S. and {Garmire}, G.~P. and {Gendreau}, K. and {Ghisellini}, G. and {Greiner}, J. and {Hill}, J.~E. and {Hunsberger}, S.~D. and {Krimm}, H.~A. and {Kulkarni}, S.~R. and {Kumar}, P. and {Lebrun}, F. and {Lloyd-Ronning}, N.~M. and {Markwardt}, C.~B. and {Mattson}, B.~J. and {Mushotzky}, R.~F. and {Norris}, J.~P. and {Osborne}, J. and {Paczynski}, B. and {Palmer}, D.~M. and {Park}, H.-S. and {Parsons}, A.~M. and {Paul}, J. and {Rees}, M.~J. and {Reynolds}, C.~S. and {Rhoads}, J.~E. and {Sasseen}, T.~P. and {Schaefer}, B.~E. and {Short}, A.~T. and {Smale}, A.~P. and {Smith}, I.~A. and {Stella}, L. and {Tagliaferri}, G. and {Takahashi}, T. and {Tashiro}, M. and {Townsley}, L.~K. and {Tueller}, J. and {Turner}, M.~J.~L. and {Vietri}, M. and {Voges}, W. and {Ward}, M.~J. and {Willingale}, R. and {Zerbi}, F.~M. and {Zhang}, W.~W.},
        title = "{The Swift Gamma-Ray Burst Mission}",
      journal = {\apj},
         year = 2004,
        month = aug,
       volume = {611},
       number = {2},
        pages = {1005-1020},
          doi = {10.1086/422091},
archivePrefix = {arXiv},
       eprint = {astro-ph/0405233},
 primaryClass = {astro-ph},
       adsurl = {https://ui.adsabs.harvard.edu/abs/2004ApJ...611.1005G}
}

@ARTICLE{Stobbart2006,
       author = {{Stobbart}, A.-M. and {Roberts}, T.~P. and {Wilms}, J.},
        title = "{XMM-Newton observations of the brightest ultraluminous X-ray sources}",
      journal = {\mnras},
         year = 2006,
        month = may,
       volume = {368},
       number = {1},
        pages = {397-413},
          doi = {10.1111/j.1365-2966.2006.10112.x},
archivePrefix = {arXiv},
       eprint = {astro-ph/0601651},
 primaryClass = {astro-ph},
       adsurl = {https://ui.adsabs.harvard.edu/abs/2006MNRAS.368..397S}
}

@ARTICLE{Brightman2022,
       author = {{Brightman}, Murray and {Kosec}, Peter and {F{\"u}rst}, Felix and {Earnshaw}, Hannah and {Heida}, Marianne and {Middleton}, Matthew J. and {Stern}, Daniel and {Walton}, Dominic J.},
        title = "{An 8.56 keV Absorption Line in the Hyperluminous X-Ray Source in NGC 4045: Ultrafast Outflow or Cyclotron Line?}",
      journal = {\apj},
         year = 2022,
        month = apr,
       volume = {929},
       number = {2},
          eid = {138},
        pages = {138},
          doi = {10.3847/1538-4357/ac5e37},
archivePrefix = {arXiv},
       eprint = {2203.11955},
 primaryClass = {astro-ph.HE},
       adsurl = {https://ui.adsabs.harvard.edu/abs/2022ApJ...929..138B}
}

@ARTICLE{gurpide2021a,
       author = {{G{\'u}rpide}, A. and {Godet}, O. and {Koliopanos}, F. and {Webb}, N. and {Olive}, J.-F.},
        title = "{Long-term X-ray spectral evolution of ultraluminous X-ray sources: implications on the accretion flow geometry and the nature of the accretor}",
      journal = {\aap},
         year = 2021,
        month = may,
       volume = {649},
          eid = {A104},
        pages = {A104},
          doi = {10.1051/0004-6361/202039572},
archivePrefix = {arXiv},
       eprint = {2102.11159},
 primaryClass = {astro-ph.HE},
       adsurl = {https://ui.adsabs.harvard.edu/abs/2021A&A...649A.104G}
}

\begin{appendix}
%\onecolumn

\section{Evaluation of secondary peaks}\label{sec:appendix_secondary}
%\textcolor{red}{CP: inserire qui plots sui scan secondari del CIE e aggiungi un paio di righe spiegando cosa c'è nella figura (vedi come ho fatto per il paper sulla NGC 55). In particolare, se ho capito bene, puoi dire che i plot secondary a $0.2c$ e rest-frame (?) persistono; piò potrebbe stare a significare che le feature fittate non sono le stesse, ovvero le componenti sono indipendenti (soprattutto per quella rest-frame che è vista praticamente in tutte le ULX in cui abbiamo almeno 3000 conteggi in RGS cfr Kosec et al. 2021, Pinto and Walton 2023). Il resto l'ho aggiunto sotto.} 
In Fig. \ref{fig:cie_scans_plus}, we show the fit improvements obtained by scanning the time-averaged spectra with the \texttt{cie} model, after including the contribution of the primary solutions (reported in Table \ref{table:physical_models} and shown in Fig. \ref{fig:CIE_results}) into the modelling together with the best-fit continuum. The solutions at velocities of about $-0.2c$ and the rest-frame ones, already observed in the previous scans, remain present, indicating that they are independent and account for different spectral features. The rest-frame component is commonly detected in ULXs with at least 3000 RGS counts \citep{Kosec2021,Pinto2023a}.

\begin{figure}[h!]
   \centering
   \includegraphics[scale=0.450]{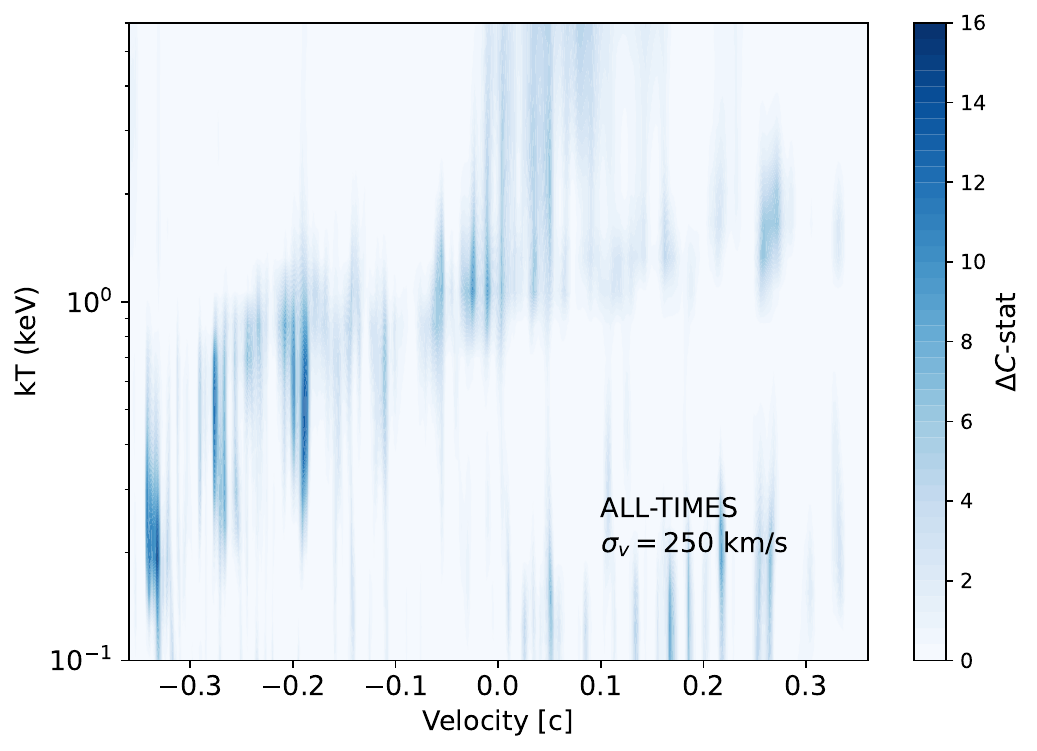}
    \vspace{-0.2cm}
   \caption{Collisional-ionisation equilibrium model scan performed on the time-averaged data, adopting a velocity dispersion of $250$ km/s, to address the significance of secondary solutions.}
    \label{fig:cie_scans_plus}
\end{figure}

We also performed simulations for the evaluation of possible secondary peaks in the kT$-v_{\mathrm{LOS}}$ maps. Assuming as template the best-fit continuum model plus three \texttt{cie} components (blueshifted, redshifted and rest-frame) for the time-averaged spectrum (see Table\,\ref{table:physical_models}), we simulated a new set of RGS and EPIC spectra using the same response files. Then we performed an identical \texttt{cie} scan for these simulated spectra with a velocity dispersion of 250 km/s. Other than the predicted three solutions ($\pm0.3c$ and rest-frame plasma) we find additional spurious solutions with lower significance ($\Delta C \lesssim 15$ or $< 2.5\sigma$) at different velocities ($0.1-0.2c$). This means that, although they are describing different features, secondary solutions with low significance (e.g. with $< 2.5\sigma$) might partially refer to spurious solutions and we should only consider as robust the results obtained with significance above $2.5 \sigma$ (ideally $>3\sigma$, see Section \ref{sec:MC_simulations}).

\section{Grids of models for photoionised plasmas}\label{sec:appendix}

In Figs. \ref{fig:pion_scans} and \ref{fig:xabs_scans}, we show the multidimensional grids for a photoionised plasma in emission and absorption, respectively, obtained for the four analysed datasets (archive-only, time-averaged, and observations 0921360101 and 0921360201) as described in Section \ref{sec:photoionisation}.

In Figure \ref{fig:pion_allexp}, we show the contribution of the most significant \texttt{pion} component in the full time-averaged spectrum.

\begin{figure}[h!]
   \centering
   \includegraphics[scale=0.45]{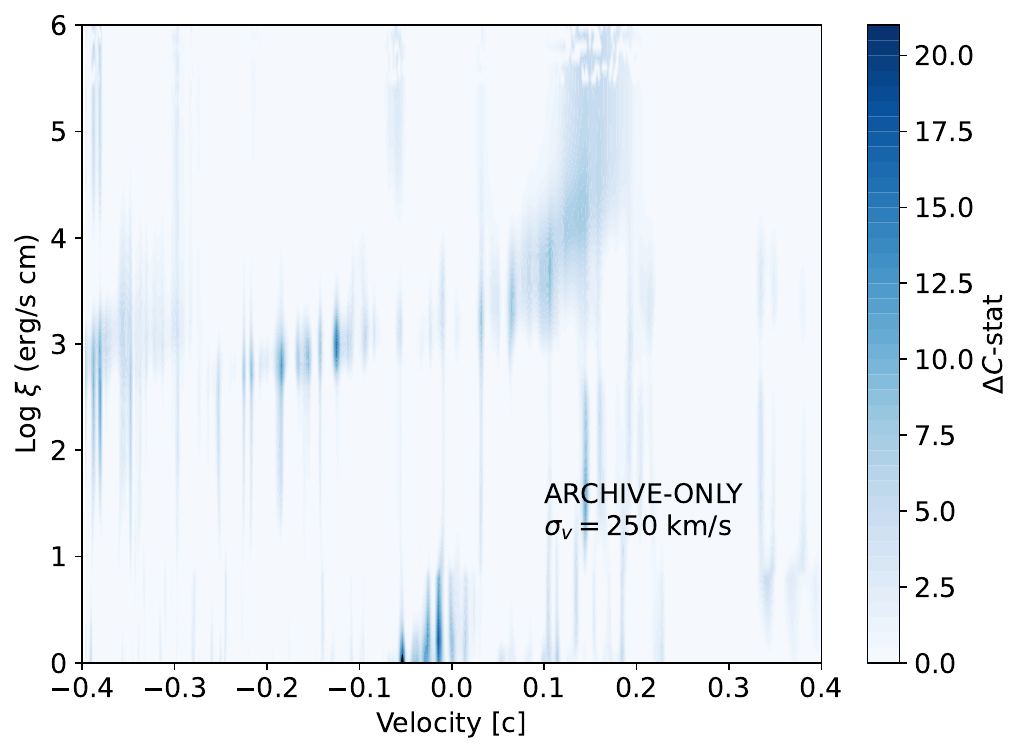}
   \includegraphics[scale=0.450]{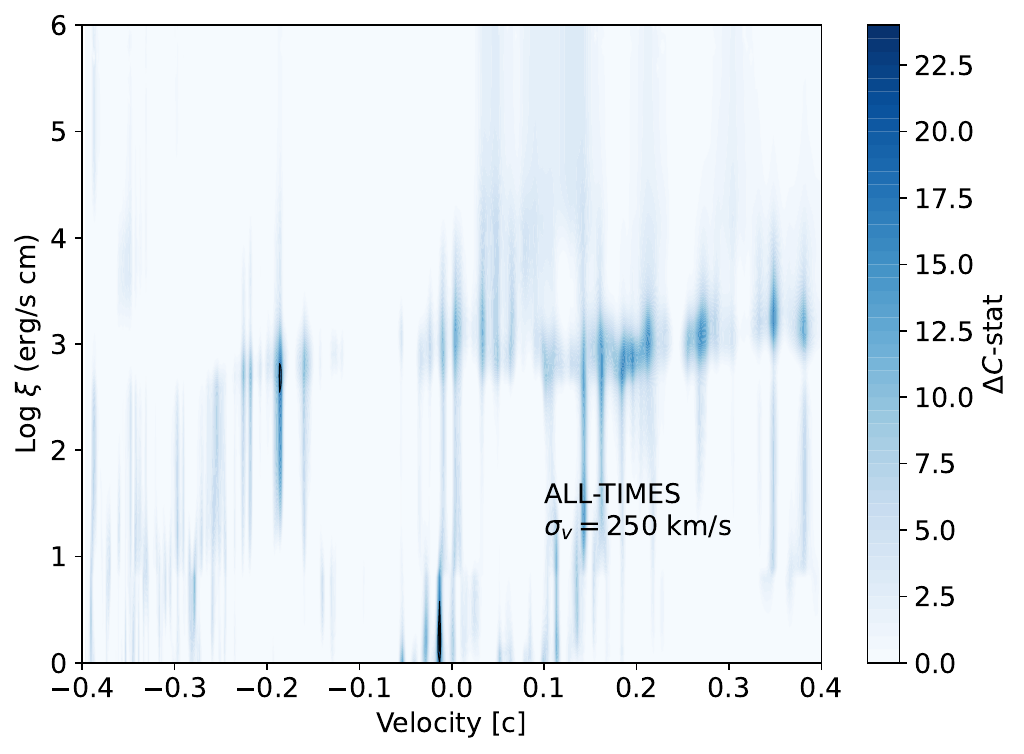}
   \includegraphics[scale=0.450]{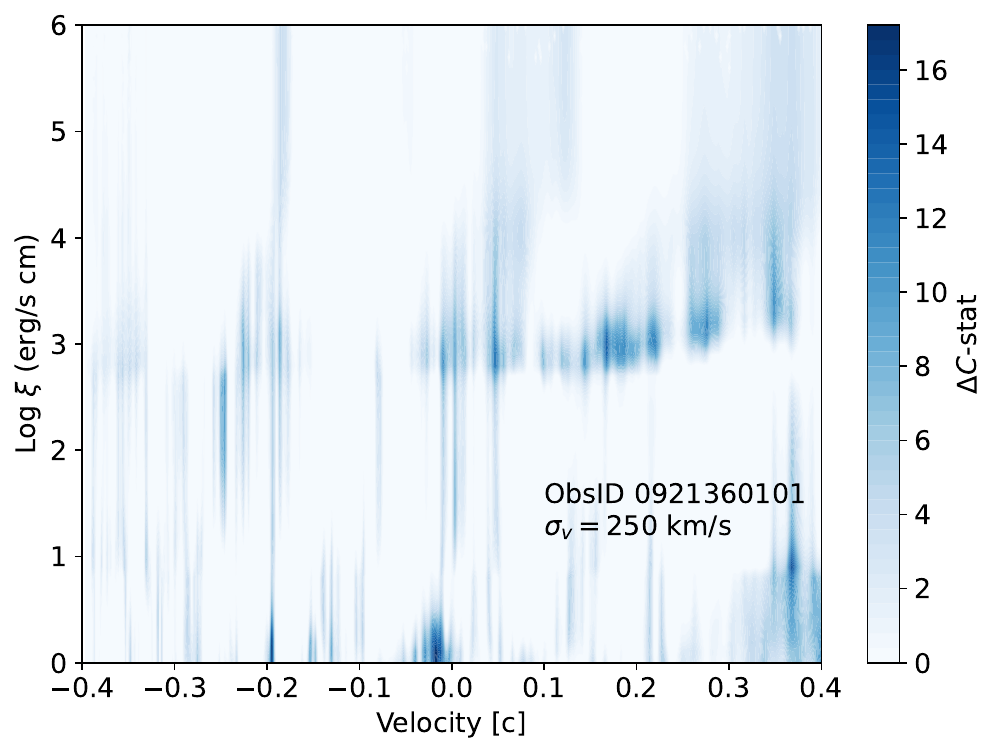}
   \includegraphics[scale=0.450]{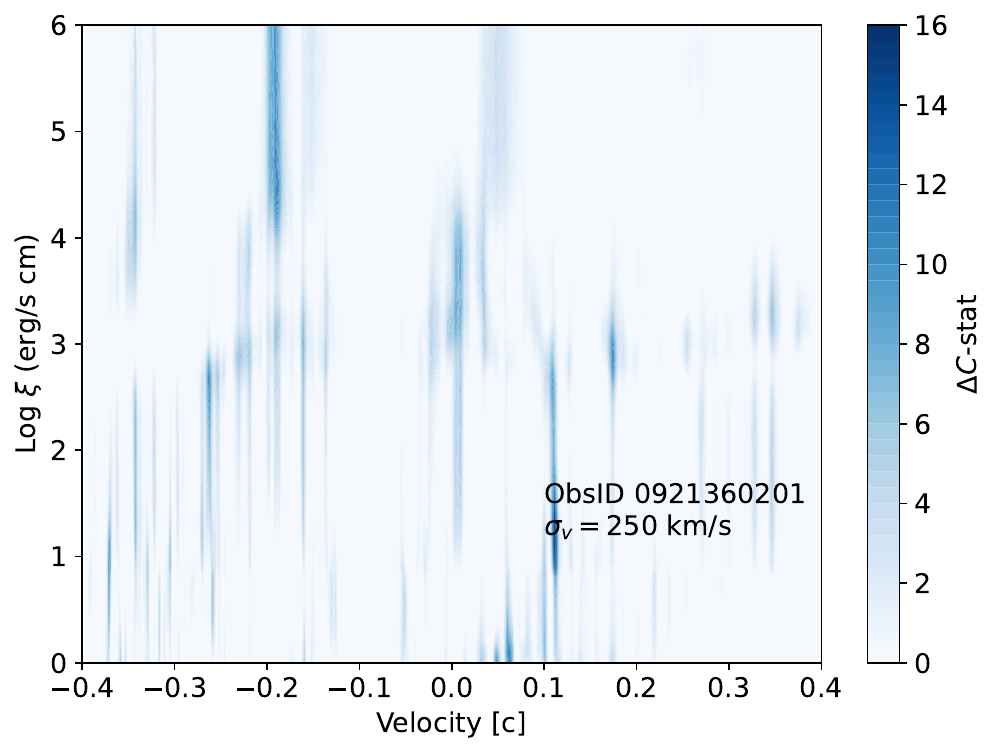}
    \vspace{-0.2cm}
   \caption{Photoionisation emission model scan for the four analysed datasets, adopting a velocity dispersion of $250$ km/s. The colours show indication of the $\Delta C$-stat improvement compared to the continuum-only model. The black contours refer to significance levels from 2.5 to 3.5 sigma with steps of $0.5\sigma$ estimated with Monte Carlo simulations (see Section \ref{sec:MC_simulations}).}
    \label{fig:pion_scans}
\end{figure}

\begin{figure*}
   \centering
   \includegraphics[scale=0.45]{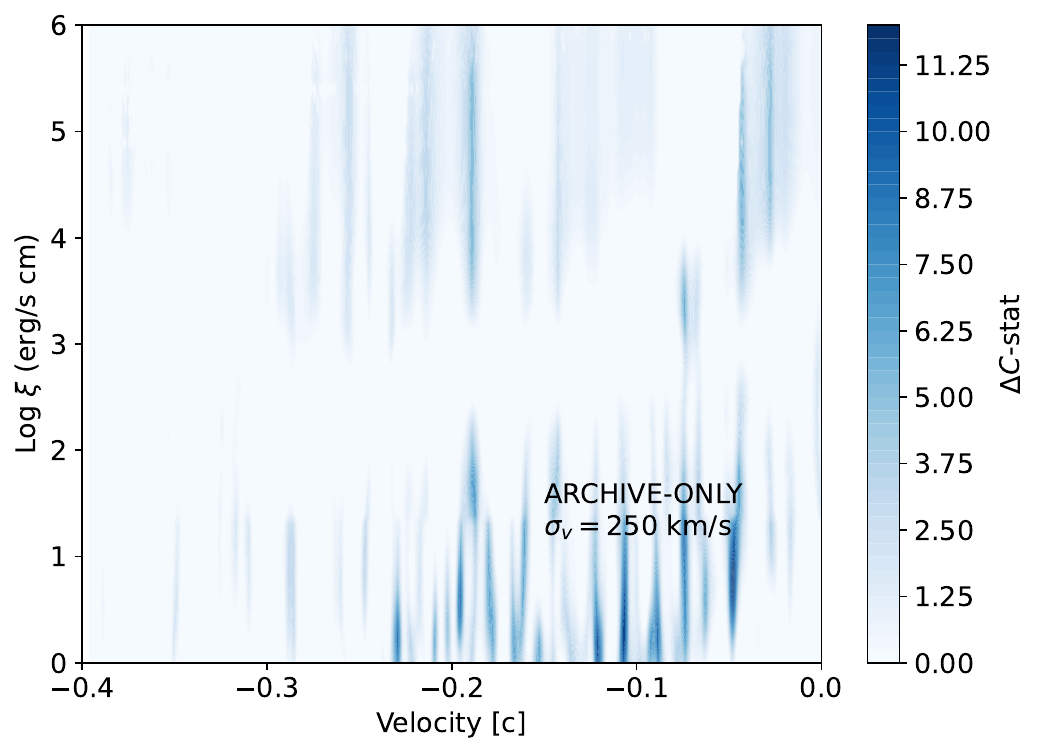}
   \hspace{0.5cm}
   \includegraphics[scale=0.45]{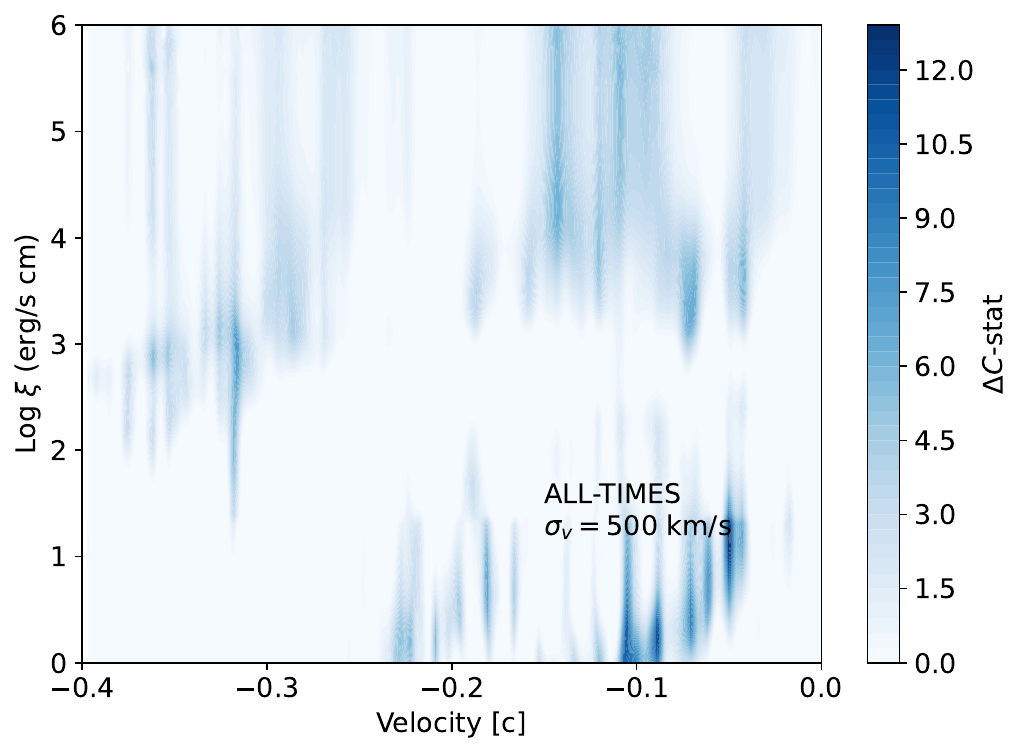}
   \includegraphics[scale=0.45]{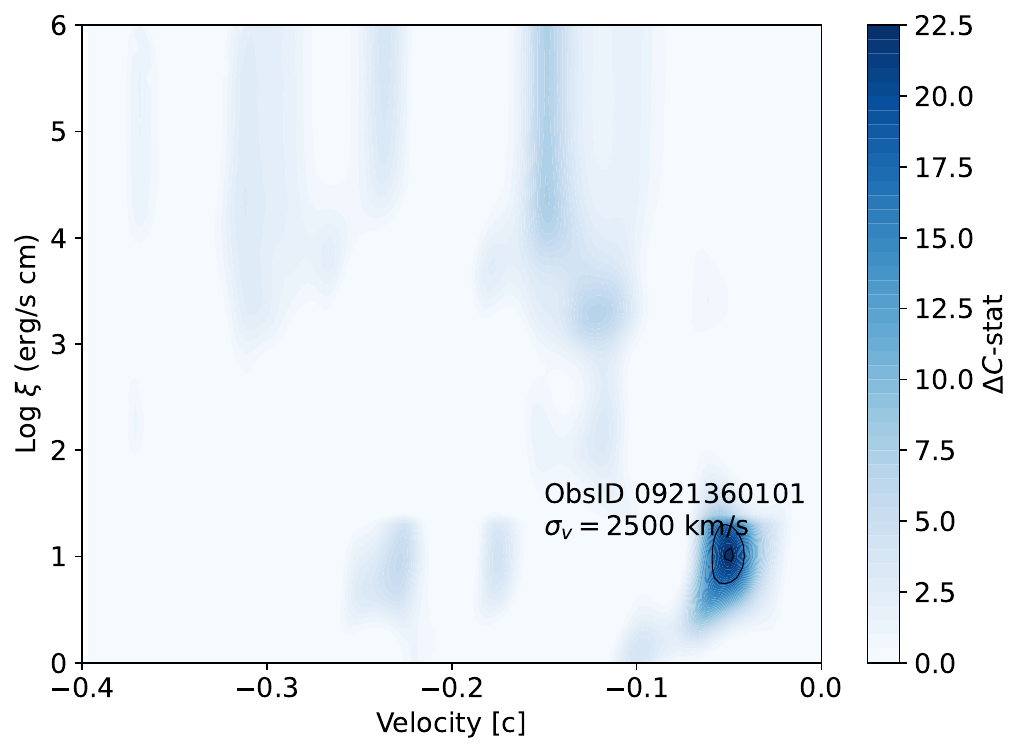}
    \hspace{0.5cm}
   \includegraphics[scale=0.45]{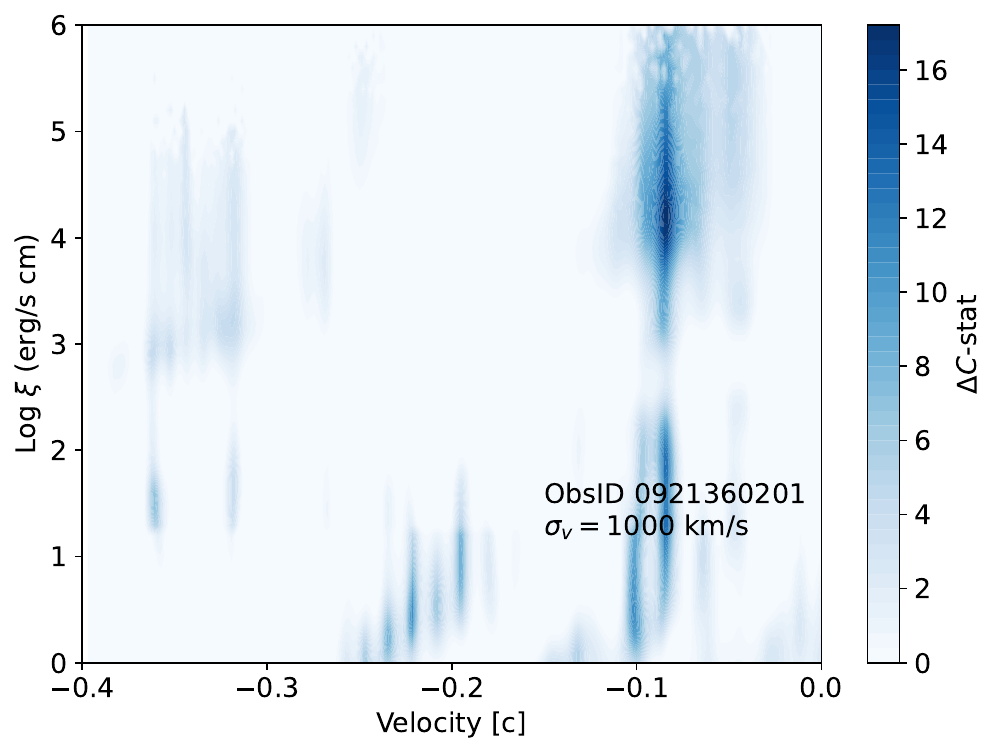}
   \vspace{-0.2cm}
   \caption{Photoionisation absorption model scan for the four analysed datasets, using the velocity dispersion that provides the best fit improvement in terms of $\Delta C$-stat. The colours show indication of the $\Delta C-$stat improvement compared to the continuum-only model. The black contours refer to significance levels from 2.5 to 3.5 sigma with steps of $0.5\sigma$ estimated with Monte Carlo simulations (see Section \ref{sec:MC_simulations}).}
    \label{fig:xabs_scans}
\end{figure*}

%\section{Contribution of spectral components}

%In Fig. \ref{fig:cie_allexp}, we show the best-fit model wich includes two \textt{cie} components (blueshifted and redshifted) and a weakly blueshifted \texttt{pion} component together with the continuum.

\begin{figure*}
   \centering
   \includegraphics[width=0.9\textwidth]{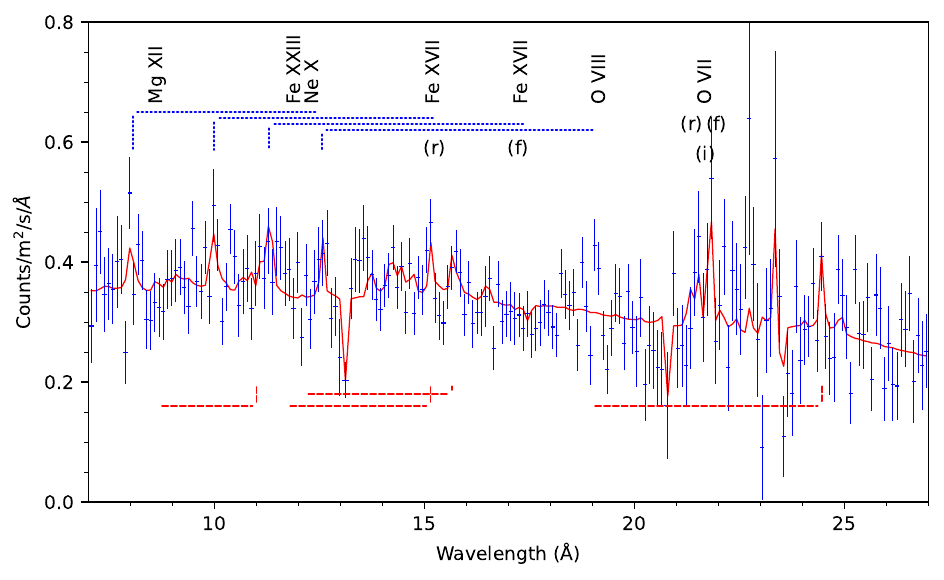}
   \vspace{-0.4cm}
   \caption{RGS spectrum in the $7-27$ \AA{} range, obtained by stacking all XMM-\textit{Newton} observations. The red curve shows the hybrid ionisation model with two fast, hotter, \texttt{cie} components and a slow, cooler, \texttt{pion} component, on top of the continuum. The rest-frame wavelengths of the most relevant lines are labelled. The dotted (dashed) lines show the velocity shift for the blueshifted (redshifted) lines. The O VII triplet is reproduced by the low-velocity \texttt{pion} component. The data are grouped by a factor of 10.}
    \label{fig:pion_allexp}
\end{figure*}

\end{appendix}

\end{document}